\documentclass[journal]{vgtc}                     

\onlineid{8346}

\vgtccategory{Evaluation}

\title{Evaluate Neighbor Search for Curve-based Vector Field Processing}

\author{%
  \authororcid{Nguyen K.\ Phan}{0009-0009-9253-7115},
  Guoning Chen
}

\authorfooter{
  \item
  	Nguyen K. Phan and Guoning Chen are with the University of Houston.
  	E-mail: 
}

\abstract{%

    Curve-based representations, particularly integral curves, are often used to represent large-scale computational fluid dynamic simulations. Processing and analyzing curve-based vector field data sets often involves searching for neighboring segments given a query point or curve segment. However, because the original flow behavior may not be fully represented by the set of integral curves and the input integral curves may not be evenly distributed in space, popular neighbor search strategies often return skewed and redundant neighboring segments. Yet, there is a lack of systematic and comprehensive research on how different configurations of neighboring segments returned by specific neighbor search strategies affect subsequent tasks. To fill this gap, this study evaluates the performance of two popular neighbor search strategies combined with different distance metrics on a point-based vector field reconstruction task and a segment saliency estimation using input integral curves. A large number of reconstruction tests and saliency calculations are conducted for the study. To characterize the configurations of neighboring segments for an effective comparison of different search strategies, a number of measures, like average neighbor distance and uniformity, are proposed. Our study leads to a few observations that partially confirm our expectations about the ideal configurations of a neighborhood while revealing additional findings that were overlooked by the community. 
}

\keywords{Curve-based vector field, neighbor search, evaluation}

\graphicspath{{figs/}{figures/}{pictures/}{images/}{./}} 

\usepackage{tabu}                      
\usepackage{booktabs}                  
\usepackage{lipsum}                    
\usepackage{mwe}                       

\usepackage{mathptmx}                  
\usepackage{float}
\usepackage[svgnames]{xcolor}

\definecolor{mygreen}{rgb}{0,0.8,0.3}

\usepackage{multirow}

\definecolor{yellowcell}{HTML}{ffd285} 
\definecolor{bluecell}{HTML}{b5d8ff}  

\usepackage{color, colortbl}
\definecolor{myred}{rgb}{1,0.3,0.3}

\definecolor{mypurple}{rgb}{0.7,0.3,0.8}

\usepackage{amsmath}
\usepackage{wrapfig}

\begin{document}

\setlength{\baselineskip}{0.97 \baselineskip}

\setlength{\abovedisplayskip}{-0pt}
\setlength{\belowdisplayskip}{-0pt}
\setlength{\abovedisplayshortskip}{3pt}
\setlength{\belowdisplayshortskip}{2pt}
\setlength{\belowcaptionskip}{1pt}
\setlength{\abovecaptionskip}{2pt}
\setlength{\textfloatsep}{9pt}
\setlength{\floatsep}{3pt}
\setlength{\intextsep}{2pt}



\firstsection{Introduction}
\label{sec:intro}

\maketitle

Curve-based representation has become popular as an intermediate representation for large-scale vector field data representation \cite{li2018data}. It is also a popular representation for visualizing and analyzing vector field data \cite{GeomSTAR09,sane2020survey}.
Although the computation and storage of curve-based representations may seem straightforward (i.e., by computing integral curves from a set of seeds and storing the list of integration points for each curve), the use of curve-based representation poses challenges to subsequent analysis and visualization tasks. This is mainly due to the incomplete representation of the original flow with the given integral curves.  
To recover the information of the flow in regions without integral curves, information from nearby curves can be drawn. To identify nearby curves (or segments on those curves), a nearest neighbor search is usually performed. Such a nearest neighbor search is often with respect to a spatial location, which is called a \emph{query point}. Consequently, we refer to such neighbor search as a \emph{point-centered} neighbor (segment) search.

Point-centered neighbor search is a well-known problem, even if the neighbors are curve segments \cite{lu2021curve}. It is usually required for information recovery in specific spatial locations in tasks like vector field reconstruction \cite{lage2006vector,bonaventura2011kernel,han2019flow} and the subsequent feature extraction \cite{FeatureBasedVF2002,gunther2018state}. Note that neighbor search with respect to a query curve segment (from an integral curve) is also useful for tasks like entropy calculation \cite{InformationTheory2010} and optimal opacity calculation for the feature-aware curve rendering \cite{gunther2013opacity,lu2021curve}. These tasks often require comparing the characteristics of an integral curve with its nearby curves. We refer to this neighbor search as \emph{curve segment centered} neighbor search. The curve segment centered neighbor search can be approximated by a point-centered neighbor search (e.g., using the middle point or the starting point of the center segment as the query point). Therefore, this paper focuses on only the point-centered neighbor search problem. 

\begin{figure}[!t]
\centering
  \includegraphics[width=0.9\linewidth]{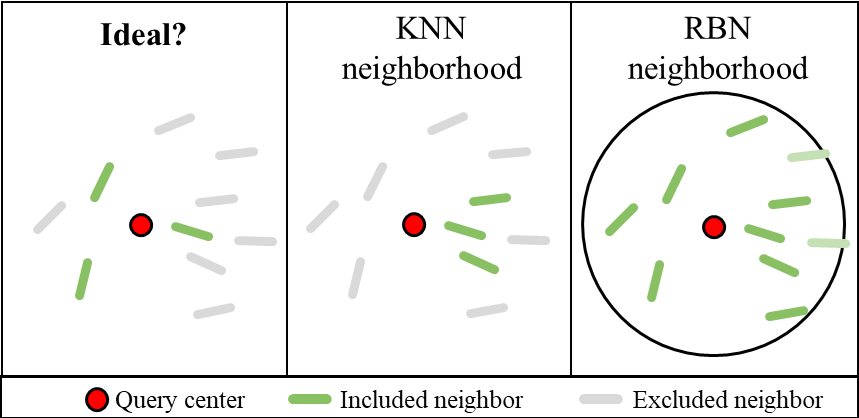}
  \caption{A 2D example of neighbor search of a given query point (red dot) from a set of candidate segments (left). KNN returns the K nearest neighbors (green) (middle),  while RBN returns neighbors falling within a sphere with a radius R, even if they are not fully enclosed by the sphere (light green). 
  }
  \label{fig:KNNissue2D}
\end{figure}

Depending on the selected neighbor search strategy, the identified nearest neighboring segments of a query point from a set of integral curves may be different. It is obvious that different neighboring segments will impact the subsequent information recovered or derived from them. Using vector field reconstruction as an example. Given a set of neighboring segments, a set of vectors can be derived based on their respective orientations. From these vectors, a vector at the query point can be interpolated (see Section \ref{sec:vfreconstruct} for details). With different neighboring segments, different vectors are used for interpolation, likely leading to different interpolated vectors. A natural question is, which set of neighboring segments will lead to an interpolated vector that is closest to the vector from the original flow at the same location? The common expectation is that ideally, the found neighboring segments should be as close to the query point as possible and they should also be evenly distributed around the query point to avoid information loss (e.g., \autoref{fig:KNNissue2D} (left)). However, such ideal configurations (especially the requirement of even distribution of segments) may not be always achievable with any neighbor search strategies given the non-evenly distributed nature of most integral curves. The questions are what are the less ideal but sufficient configurations (i.e., how close is close enough, and how even is good enough?), and which search strategy is most likely to find neighbors with such less ideal but sufficient configurations?
Answering these questions is not trivial, as such less ideal but sufficient configurations may vary at different locations of the flow. 
A more holistic and statistical approach is needed in order to draw conclusions from a large set of samples. This motivates our study.

Our study is the first comprehensive study to assess the impact of different neighboring segments returned by different neighbor search strategies on tasks related to curve-based vector field processing.
In our study, we focus on streamline data sets. We select the popular KNN and RBN neighbor search methods (\autoref{fig:KNNissue2D}) and a variety of distance metrics, including the shortest, longest, and average distance, for our study. This is the first time different distance metrics are considered for such a study. 
We consider two tasks that require neighbor search, i.e., vector field reconstruction (\autoref{sec:vfreconstruct}) and saliency calculation for individual segment (\autoref{sec:predictedsaliency}). The former may be needed for analysis, while the latter is often used to emphasize portions of the curves that are of interest in visualization. In addition, the former performs neighbor search centered at locations that need not be on the input curves, while the latter search neighbors for the individual segments of the input curves. They represent different scenarios of neighbor search for curve-based vector field processing. We wish to clarify that our goal is not to improve the practice of vector field reconstruction and saliency calculation.



To characterize the configurations of neighboring segments, we propose a number of measures (\autoref{sec:neighbormeasure}), including the average distance of the neighboring segments to the query point and the uniformity of the spatial coverage of the segments around the query point. These measures fill the gap of the current lack of such measures for characterizing neighboring segment configurations.

To evaluate the impact of different search methods and distance metrics, we exhaust their possible combinations (i.e., KNN+shortest distance, KNN+longest distance, and so on). 
For each combination, we utilize it to identify the neighboring segments for every grid point, which are used to perform vector field reconstruction and saliency calculation.
In addition, for each combination, we set different search parameters (e.g., different $K$ values for KNN and different radius $R$ for RBN). For each representative flow, we derive different sets of streamlines using a different number of seeds and placement strategies (e.g., feature-aware, uniform, and controlled randomization). All these variations and combinations result in thousands of tests, ensuring the generality of our study.

Analyzing this large number of results to identify meaningful findings is challenging. We resort to statistical analysis coupled with conventional visual inspection to analyze these results. 
For each test, we first inspect the average performance (e.g., reconstruction error) across all query points, followed by a closer look at the individual query points and their respective neighborhood grouped by the proposed average distance and uniformity metrics statistically.
Since it is impossible to compare the KNN-based result with a specific K with the RBN result using a particular R, to be fair we opt to compare the best KNN result with the best RBN result for each input streamline dataset in the two respective tasks. Our systematic analysis of the experiment results lead to several important observations. While some are aligned with our expectations, many are not.  To better understand this, we subdivide the result of each streamline data set into subgroups for a detailed statistical analysis. This further reveals the reasons behind the different performances of KNN and RBN under different streamline configurations. 


Even though the setup of our study is still limited, we were able to reveal some complex relations between the configurations of the input streamlines and their impact on finding the appropriate sets of neighbors for the two tasks. Our statistical analysis on the level-of-detail and multi-conditioning grouping of neighborhoods may help explain some of our findings.
Furthermore, our study shows the effectiveness of the proposed measures for characterizing the configuration of neighboring segments, which can be used to guide the development of future neighbor search methods. We also believe that the setup of the experiments and the analysis of the results used in our study can be adapted for other similar studies for the development of effective algorithms for the processing of curve-based vector fields in the future.

\section{Related Work}
\label{sec:related_work}

Integral curves are commonly used to depict the geometric behavior of various flow data \cite{GeomSTAR09,STARSurfVis12}. They are one of the most commonly used representations of 3D vector fields. 
In the following, we briefly review the most important and relevant works for integral curve generation and processing, along with the relevant neighbor search techniques. 

\paragraph{Integral curve computation.}
To obtain an effective integral curve representation, a proper seeding strategy is required.
Depending on different goals, different seeding strategies can lead to evenly-spaced streamlines in either the physical (object) space \cite{jobard1997creating,liu2006advanced,mebarki2005farthest,chen2007similarity}  or the image space \cite{turk1996image,spencer2009evenly,li2007image}. Other strategies can produce feature-aware streamlines \cite{verma2000flow,chen07,wu2009topology} and streamlines that best represent the flow based on information theory \cite{FlowInfo10}. A recent survey paper \cite{sane2020survey} provides a detailed review of existing integral curve seeding and placement strategies. To handle the placement of integral curves for large-scale flow data, many parallel and distributed computing strategies have been proposed \cite{zhang2018survey}.



\paragraph{Integral curve processing.}
A few tasks can be performed with the given integral curves. 
For example, the construction of the original vector fields from the given set of integral curves may be performed for subsequent analysis. 
There are only a few works on reconstructing vector fields from the curve-based data sets \cite{han2019flow,IAVFSR21}.
The reconstructed vector field can be used to extract features that are hard to extract from the given curves, such as the topological features \cite{LAR07}, flow separation structure \cite{SHADDEN2005271,pobitzer2011state}, and other physics-relevant features (e.g., vortices \cite{gunther2018state}). This is because the extraction of these features requires a continuous representation of the vector fields. There currently exists little work in the extraction of meaningful and accurate features \emph{directly} from the input integral curves. 

\paragraph{Integral curve-based visualization and exploration.}
To visualize the computed integral curves for an effective analysis of the flow behaviors, different rendering and interaction techniques have been proposed. 
G\"unther et al. \cite{gunther2013opacity} proposed an optimization framework to determine the optimal opacity values for the individual curve segments to reveal the inner patterns of interest that may be occluded due to the dense placement of 3D integral curves. 
Tong et al. \cite{tong2015view} introduced a view-dependent streamline deformation method to allow the user to unveil hidden structures and patterns. 
Recently, Lu et al. \cite{lu2021curve} introduced an effective KD-tree data structure for the individual curve segments decomposed from the input integral curves to support an interactive exploration and query of 3D curves. Viewpoint selection techniques \cite{LeeMSC11,tao2012unified} are also used to automatically select viewpoints for streamline rendering to better depict flow patterns.
To reduce the clutter and occlusion issues caused by the dense 3D integral curves, integral curve clustering \cite{shi2021integral} or bundling \cite{Yu2012HSB} can also be performed to aid the selection of representative integral curves to render.
Recently, Nguyen et al.~\cite{TAC2019,nguyen2021physics} introduced a framework for the exploration of unsteady vector fields through the characterization of pathlines behavior. 

\paragraph{Nearest neighbor search.} 
Nearest neighbors of a curve or curve segment are usually identified to support curve selection \cite{sane2020survey,moberts2005evaluation}, segmentation  \cite{wang2014pattern}, and abstraction  \cite{everts2015exploration}. Searching for nearest neighbors with many curve segments can be slow. Hence, to find the (approximate) nearest neighbor faster, a few algorithms \cite{hoel1992qualitative,elseberg2012comparison} have been proposed. Among them, KD-trees \cite{arya1998optimal} and their variants are the most popular approaches. A KD-tree is constructed by recursively partitioning the space with axis-aligned planes. With this partitioning and its hierarchical spatial relation, KD-tree supports an efficient nearest neighbor search in logarithmic time. To further improve the query performance of KD-tree, efforts have been made to optimize the KD-tree structure \cite{goldsmith1987automatic,lu2021curve}. In our experiment, we use KD-tree to accelerate KNN and RBN.
\section{Point-based Neighbor Search}
\label{sec:pbneighborsearcn}

Point-centered neighbor search aims to identify line segments, derived from 3D curves, in the vicinity of a specific query point $\mathbf{p}$. 

Given an integral curve with $N$ integration points, we sub-divide it into $N-1$ line segments (i.e., two consecutive integration points form a segment). This decomposition leads to individual straight-line segments, simplifying our subsequent processing (e.g., distance measuring and neighborhood construction). Note that other more sophisticated decomposition strategies, like the curvature-based decomposition \cite{lu2021curve}, may introduce additional variables to our study (e.g., curvature threshold for decomposition), thus, we leave them for future studies.

Two prevalent search strategies are employed in this context: K-nearest neighbor search (KNN) and radius-based neighbor search (RBN).
For KNN, segments are arranged in ascending order of their distance to $\mathbf{p}$ using the distance measure $\mathbf{d}(\mathbf{p}, L_i)$ and the top $K$ segments are chosen as neighbors. In contrast, RBN identifies neighbors within a sphere of radius $r$ centered at $\mathbf{p}$, where segments with distances less than $r$ are considered neighbors.

In our preliminary experiments, both default KNN and RBN exhibited sub-optimal performance near domain boundaries and in regions with sparse distribution of curves, necessitating modifications to improve their accuracy and reliability.

\textbf{RBN Modification:} For data sets with substantial sparse regions, our initial observations revealed that RBN often fails to locate any neighboring segment within the radius $R$ when performed on grid points situated in these sparse areas. Although a common solution is to increment $R$ slightly, we found this problematic. When $R$ is already large, even minor increments drastically expand the search area. This often results in the identification of an excessive number of neighbors. Instead of merely increasing $R$, we adopted an alternative strategy. When no neighboring segments are located within $R$, we simply select the nearest segment. While this approach effectively mitigates the error surge in sparse regions due to RBN overfitting, it potentially provides RBN an undue advantage over KNN. In essence, for sparse regions, RBN is replaced by a KNN with $K=1$.

\textbf{KNN Modification:} For KNN, we noticed elevated errors near domain boundaries, particularly at corners (\autoref{fig:volrendC}). Utilizing a consistent value of $K$ for the entire data set forces the algorithm to stretch further to find the same number of $K$ neighbors in sparse or boundary regions. This, in turn, escalates the error, given the increased average distance to the neighborhood. Hence we introduced a modification to KNN. For each query point, we calculate the distance, $d_{corner}$, to the nearest corner. This distance serves as an exclusion threshold. Neighbors identified by KNN are excluded if their distance to the query point exceeds $d_{corner}$. If no segments meet this criterion, only the nearest neighbor is retained.

Next, we select the distance metric to determine the neighbor relations between a point and a line segment. Three distance metrics are considered in our study: 

\begin{itemize}[noitemsep,nolistsep]
    \item \textbf{Shortest Distance ($\mathbf{d}_{\min}$)}: a straightforward to compute since $L_i$ is a straight segment based on our decomposition. Given by 
    \[\mathbf{d}_{\min}(\mathbf{p}, L_i) = \min(d(\mathbf{p}, \mathbf{q})) | \forall \mathbf{q}\in L_i,\]
    
    \item \textbf{Longest Distance ($\mathbf{d}_{\max}$)}: the greater of the distances between $\mathbf{p}$ and the two endpoints of $L_i$. Represented as
    \[\mathbf{d}_{\max}(\mathbf{p}, L_i),\]
    
    \item \textbf{Average Distance ($\mathbf{d}_{mean}$)}: approximates the mean of the shortest and longest distances. Denoted by 
    \[\mathbf{d}_{mean}(\mathbf{p}, L_i) = \frac{1}{2}(\mathbf{d}_{\min} + \mathbf{d}_{\max}),\]
\end{itemize}


\begin{figure}[!t]
\centering
  \includegraphics[width=0.85\columnwidth]{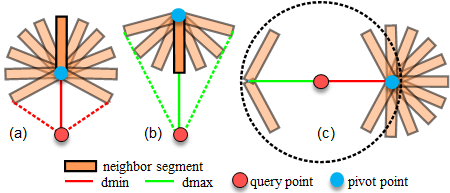}
  \caption{Illustrating how a neighbor segment with the same $\mathbf{d}_{min}$ value to a grid point can yield very different $d_{max}$ values and vice-versa (case (a) and (b)). Case c illustrates how with the same R radius parameter, the number of possible neighbor segments accepted by using $\mathbf{d}_{min}$ is significantly higher than those that will be accepted based on $\mathbf{d}_{max}$. For (c), the dashed circle indicates a search area limited by a radius R where neighbor segments whose distance exceeds R are excluded.
  }
  \label{fig:pointlinecase}
\end{figure}

Although the shortest distance is fundamental, it might not fully capture segment-to-point positional nuances. The average and longest distances can partially address this, but come with their own challenges, such as potential inaccuracy or excluding potential neighboring segments. 

\autoref{fig:pointlinecase} demonstrates the variability in neighboring segment selection based on the distance metric used, given the same grid point and neighbor search strategy. In \autoref{fig:pointlinecase}a, we observe a variety of segment orientations that produce identical values for $\mathbf{d}_{\text{min}}$, but result in significantly different $\mathbf{d}_{\text{max}}$ values. Conversely, \autoref{fig:pointlinecase}b presents situations where the $\mathbf{d}_{\text{max}}$ remains constant, but the corresponding $ \mathbf{d}_{\text{min}} $ varies widely.
This disparity implies that the set of neighboring segments identified using $ \mathbf{d}_{\text{min}} $ might differ from the set obtained with $ \mathbf{d}_{\text{max}} $. Furthermore, segments determined by $ \mathbf{d}_{\text{max}} $ often align more closely with the grid point, resulting in segment portions that are nearer to this point. In contrast, $ \mathbf{d}_{\text{min}} $ can select segments oriented away from the grid point, potentially capturing segments that are both oriented further away and are more distant along the segment length relative to the grid point. \autoref{fig:pointlinecase}c seeks to visually represent these behavioral differences between $ \mathbf{d}_{\text{max}} $ and $ \mathbf{d}_{\text{min}} $ given a fixed radius value $ R $. 

\section{Neighbor Segment Configuration Characterization}
\label{sec:neighbormeasure}

In this section, we introduce a number of measures for the characterization of different configurations of neighbor segments returned by a specific neighbor search method for the point-centered query. 
These measures will be associated with the quality of the subsequent vector field reconstruction task, which helps us assess the impact of different search strategies on this task and helps us identify possible ideal configurations of neighbor segments.

To characterize the configuration of a set of neighbor segments, we focus on two aspects: (1) the closeness of segments to the query point or query curve and (2) the uniformity of their distribution in space and around the query point or query curve. These two aspects correspond to the two intuitions mentioned in Section 1.




\subsection{Average Distance of Segments}
\label{sec:avgDistance}

Given a set of segments $L$ (e.g., the nearest neighboring segments of a query point returned by KNN or RBN), we characterize the closeness of all segments $L_i \in L$ to the query point $\mathbf{p}$ as their average distance to $\mathbf{p}$. That is, $\frac{1}{|L|} \sum_{i} \mathbf{d}(\mathbf{p}, L_i)$ for point-center neighbors. Here, $\mathbf{d}(\mathbf{p}, L_i)$ is the distance metric (i.e., shortest, longest, or average distance) used to identify the neighboring segments.

\subsection{Uniformity of Spatial Coverage}
\label{sec:uniformitymeasure}

\begin{wrapfigure}{r}{0.25\columnwidth}
\centering
  \includegraphics[width=0.25\columnwidth]{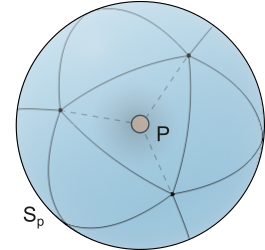}
  \label{fig:neighborhood}
\end{wrapfigure}
The spatial coverage uniformity metric we employ evaluates the even distribution of neighboring segments around a given query point. 
To measure the uniformity of the neighboring segments found around $ \mathbf{p} $, 
we first construct a sphere centered at $ \mathbf{p} $, denoted as $ S_{\mathbf{p}} $ 
(see the inset to the right).
The sphere's radius $ r $ is selected such that all identified neighbors, whether located through KNN or RBN methods, reside within $ S_{\mathbf{p}} $.

To achieve a consistent subdivision of this space and optimize computational efficiency, we use a spherical icosahedron inscribed within $ S_{\mathbf{p}} $. Each face of the icosahedron serves as the base for a three-dimensional spatial bin that extends inward to the center $ \mathbf{p} $. Given that an icosahedron has 20 faces, this results in 20 spatial bins. Each bin is uniquely associated with one of the icosahedron's faces and extends throughout the volume of $ S_{\mathbf{p}} $. 

To further reduce computation costs, we implement a 2D lookup table based on the projected spherical coordinates. The table is constructed by projecting each cell from the 2D spherical coordinate space onto the 3D space to determine which bin it aligns with, subsequently marking that cell with the appropriate bin index. Our approach has demonstrated robust accuracy, with the lookup table yielding 99.6\% precision in uniform points testing. Furthermore, this implementation has achieved an impressive speed increase, more than 10 times faster than the manual spatial bin lookup in 3D space. Note that other triangulation of the sphere $ S_{\mathbf{p}} $ with uniform-size triangles can be used here to achieve different numbers of bins. However, in our study, we found that 20 bins are sufficient in most cases.

Given the above $S_{\mathbf{p}}$, we flag the bins that intersect with at least one neighbor segment. Let us assume $M$ bins are flagged out of $N$ total bins. The uniformity measure is then simply defined as $\mu=\frac{M}{N}$. The closer to 1 is $\mu$, the more uniform (or even) the distribution of these neighbors around $\mathbf{p}$ (or $\mathbf{c}$). 

Note that the above simple strategy does not consider the fact that different bins may be occupied by different numbers of segments. In addition, even if a bin is flagged, it may only contain a small portion of a segment. To counter this, one can use a weighing strategy for each flagged bin based on how much a segment falls within it and how many segments intersect with it. Nonetheless, in this study, we opt for a simple measure and leave the more comprehensive one for future work.



\section{Vector Field Reconstruction}
\label{sec:vfreconstruct}

First, we use vector field reconstruction to evaluate the impact of different neighbor search strategies. 
This is because it allows us to quantitatively measure the impact through the calculation of reconstruction error. 
Given a set of streamlines, we will reconstruct the vector values at the individual grid points of the same spatial discretization used to store the original flow. These reconstructed vector values will be compared against the ground truth values stored in the original flow.

To reconstruct the velocity vector at a grid point, we need to first locate a number of neighboring segments surrounding it and then extrapolate the velocity vector based on the orientations of these segments. Based on the found neighboring segments, we employ a standard distance-based weight sum to extrapolate a velocity vector. Assuming $N$ represents the set of neighbor segments found for a grid point $\mathbf{p_j}$, the reconstructed vector is computed as follows.
\begin{equation}
    V\left(\mathbf{p_j}\right)=\sum _{i=1}^{\left|N\right|}\:w_iV\left(N\left(i\right)\right); \ \ \ \ 
    \label{eq:vfreconstruct}
\end{equation}
\noindent where $V(N(i))$ is the vector derived from the segment $N(i)$, and $w_i$ is the weight obtained from a chosen distance weighing function performed on segment $N(i)$ \cite{Yao2013ComparisonOF}.

The goal is to reduce the difference (or error) between the reconstructed vector field and the original vector field. To measure the reconstruction error, we compute the mean error, which is defined as $e_{mean} = \frac{1}{M} \sum_{i=0}^{M-1} e(V(\mathbf{p_i}), V_g(\mathbf{p_i}))$, where $e(V(\mathbf{p_i}), V_g(\mathbf{p_i}))$ measures the error between the reconstructed vector $V$ and the ground truth vector $V_g$ at grid point $\mathbf{p_i}$. Here, we use $e(V(\mathbf{p_i}), V_g(\mathbf{p_i}) = ||V(\mathbf{p_i}) - V_g(\mathbf{p_i})||$. This error magnitude encodes both the angle and magnitude differences between $V$ and $V_g$.
The normalized error, $\frac{||V(\mathbf{p_i}) - V_g(\mathbf{p_i})||}{||V_g(\mathbf{p_i})||}$, can also be used. However, in our study, we found it usually leads to much larger values for places with small velocity magnitude, dictating the average error. Thus, we do not use the normalized error.
\paragraph{Interpolation method decision}
To identify the most effective weighing function for our reconstruction task, we carried out initial tests on all seven data sets using three distinct weighting schemes. We present the preliminary results of several interpolation methods:

\begin{enumerate} [noitemsep,nolistsep]
    \item \textbf{Uniform Weighting}: We begin by evaluating the results without employing any weights. This served as our baseline, enabling us to discern the improvements that various interpolation schemes could offer over it.
    \item \textbf{Gaussian Weighting}: We explored the Gaussian weighting approach, an example of a radius-based fall-off function (RBF), where each neighboring segment is accorded a weight derived from the Gaussian function. The formula for Gaussian weighting is given by:
   $ w(x) = e^{-\frac{x^2}{2\sigma^2}} $
    In our experiments, the optimal value of $ \sigma $ was determined to maximize the efficacy of the interpolation.
    \item \textbf{Inverse Distance Weighting}: Lastly, we adopted the inverse distance weighting method for interpolation. 
    $w_i=\frac{\frac{1}{d_i^2}}{\sum _{i=1}^{\left|S\right|}\frac{1}{d_i^2}\:} $
\end{enumerate}


Our experiments show that inverse distance weighting outperformed the other schemes. The fixed length of the segments in our streamline data sets, designed to prevent segments from approaching zero length near fixed points, might have compromised the efficacy of the Gaussian weighting. This potential limitation of Gaussian weighting led us to \textbf{select inverse distance weighing} for the subsequent evaluation.

In addition to the above three interpolation strategies, we also tested another approach for our reconstruction tests. Instead of extracting a single vector from each neighboring segment, we utilized the \emph{central-difference vector}, which is achieved by averaging the forward and backward streamline segments of each neighboring segment. The outcome of this method, interestingly, was marginally inferior compared to using the original neighbor segment's vector. Therefore, we decided to use the vectors corresponding to the individual line segments for reconstruction.






\section{Saliency Calculation}
\label{sec:predictedsaliency}

Second, we calculate a saliency value for each segment based on the difference of the orientations of its neighboring segment with respect to it. This measure aims to identify segments in interesting flow regions that are characterized by large changes of flow directions (e.g., flow separation and vortex cores). 


\subsection{Saliency Approximation for Segments}
\label{subsec:predictedsaliency}

Given a segment $L$ with orientation $\mathbf{v}$, we find K neighboring segments based on their distance to the starting point $\mathbf{p}$ of $L$ using either KNN or RBN. A saliency value for $L$ can be computed as follows \cite{Phan2022DirectNeighbor}.
\begin{equation}
S_{cal}(\mathbf{p}) = \frac{\sum_{i=1}^K w_i \arccos\left(\frac{\mathbf{v} \cdot \mathbf{v_i}}{|\mathbf{v}||\mathbf{v_i}|}\right)}{\sum_{i=1}^K w_i}
\end{equation}

\noindent where $\mathbf{v_i}$ is the orientation of the $i$-th neighboring segment and $w_i$ is its weight, calculated using inverse distance weighting
$w_i = \frac{1}{d(\mathbf{p}, \mathbf{p_i})^2}$. 
\noindent $d(\mathbf{p}, \mathbf{p_i})$ is the distance between the query point $\mathbf{p}$ and the start point of the $i$-th neighboring segment $\mathbf{p_i}$. $\arccos\left(\frac{\mathbf{v} \cdot \mathbf{v_i}}{|\mathbf{v}||\mathbf{v_i}|}\right)$ computes the angle between vectors $\mathbf{v}$ and $\mathbf{v_i}$.





To evaluate how good the calculated saliency value for each segment is, we need to have a reference saliency. Different from the vector field reconstruction task where the ground truth is the original flow, there is no ground truth saliency value for a segment. To remedy that, given a segment $L$ with a starting point $\mathbf{p}$ and an orientation $\mathbf{v}$, we sample $N$ points uniformly on a circle centered at $\mathbf{p}$ with a radius $R$ on a plane perpendicular to $\mathbf{v}$. For each sample point $\mathbf{s_i}$, we compute the vector $\mathbf{v_i}$ from the original vector field. Then, we compute a \emph{reference} saliency value for $L$ as follows.
\begin{equation}
S_{ref}(\mathbf{p}) = \frac{1}{N} \sum_{i=1}^N \arccos\left(\frac{\mathbf{v} \cdot \mathbf{v_i}}{|\mathbf{v}||\mathbf{v_i}|}\right)
\end{equation}

With the above saliency calculation, the choice of a proper $R$ needs to be decided. As the saliency is expected to capture the variation in the streamline orientation that may indicate features, we rely on visual inspection and our prior knowledge of the flow to choose an $R$ that leads to the saliency values of the segments better highlighted the known structures. 


With the reference saliency, the quality of the calculated saliency may be evaluated via MSE (similar to the vector field reconstruction task). However, we notice that in most cases, the calculated saliency value is much lower than the reference saliency.

\section{Data sets and Setup of our Study}
\label{sec:setup}

\begin{figure}[!t]
\centering
  \includegraphics[width=0.95\columnwidth]{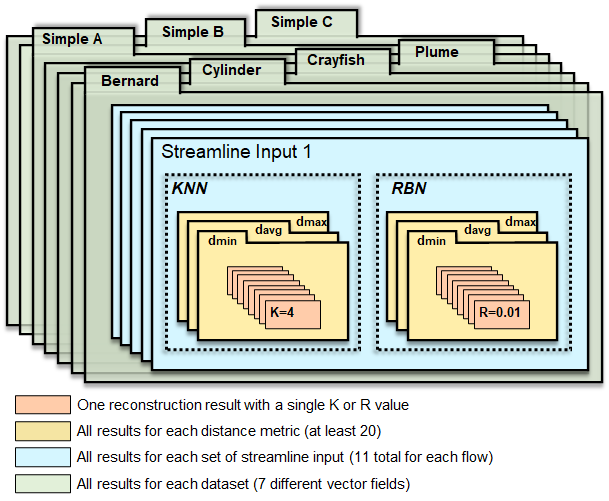}
  \caption{The organization of the reconstruction tests based on different flow data sets, different streamline placement strategies, and different neighbor search strategies combined with different distance metrics. For each search strategy, different parameter values are used to search for the best reconstruction results. This demonstrates the complexity of our study when considering different factors.}
  \label{fig:challenges}
\end{figure}

We utilized seven vector field data sets for our study. These data sets encompass three formula-based analytic flows (whose formulas can be found in the supplemental document), the Bernard convection flow, the flow behind a cylinder, the plume, and the crayfish. The latter four data sets were also used in a recent study \cite{shi2021integral}. Streamlines for these data sets were all computed with RK4 integration and fixed step sizes. 

Owing to the extensive number of tests conducted across various streamline sets for each data set—totaling over \textbf{4228}  tests—we introduced Figure \ref{fig:challenges} to delineate our organization of the tests. This organization spans four levels: vector field data set, streamline input data set, distance metric, and finally, individual tests based on the K parameter for KNN and the search radius R for RBN.


To identify the neighbor segments for reconstruction and saliecy calculation, we use the KNN and RBN neighbor search strategies, respectively. They are combined with three different distance metrics described in Section \ref{sec:pbneighborsearcn}. For KNN, we vary the value of K (i.e., 4, 6, 8, ...). For RBN, we vary the search radius $R$. However, different from KNN, the proper search radius highly depends on the complexity of the individual flow and the spatial density of the provided streamlines. To combat that, we use slightly different sets of $R$ values for different data sets so that their results are comparable to the KNN-based results.

Note that our goal of varying the values of $K$ and $R$ is to identify the best $K$ and $R$ for each data set for each task. To achieve that, different data sets may require different numbers of tests. On average, each streamline data set requires at least 6 different values of $K$ and 6 different values of $R$, leading to more than \textbf{4228} tests that we conducted for this study. 

\paragraph{Varying seeding strategies.}
To ensure robustness and consistency in our evaluation results, we produced multiple streamlines data sets for each of the seven vector fields. Every vector field data set comprises three distinct "levels" of seed numbers. For each level, a specified number of seeding points, $S_{N}$, is derived from the first strategy - uniform seeding. For the second strategy, a parallel streamline data set is created using $S_{N}$, initially uniformly distributed, but each seeding point is then "jittered" by an amount equivalent to 90\% of the spacing length. Yet another streamline seeding data set is derived from $S_{N}$, but the seeds are placed entirely at random. During our evaluation, we compared the three streamlines data sets for each $S_{N}$ from every vector field data set. 

Beyond the strategy of uniformly placed seeding, we generated a feature-aware streamlines data set for all seven flows. To formulate these feature-aware streamlines, we took the following approach for each vector field data set:
\begin{enumerate}[noitemsep,nolistsep]
\item Generate uniformly spaced seeding points.
\item Apply Simulated Annealing to ``nudge'' each seeding point towards the more ``interesting'' regions of the vector field. We determined the ``interest'' of a region using the gradient field, where a higher gradient magnitude typically signifies more activity within the vector field.
\item Compute streamlines from the modified seeding points.
\end{enumerate}

The inclusion of streamlines with different seeding and placement strategies aimed to assess consistency in results.



\section{Results and Analysis}
\label{sec:results}
In this section, we present our findings in a structured, top-down manner. Our aim is to provide a comprehensive view of the performance patterns of different search methods that we observed. 

\begin{figure}[!t]
\centering
  \includegraphics[width=0.93\columnwidth]{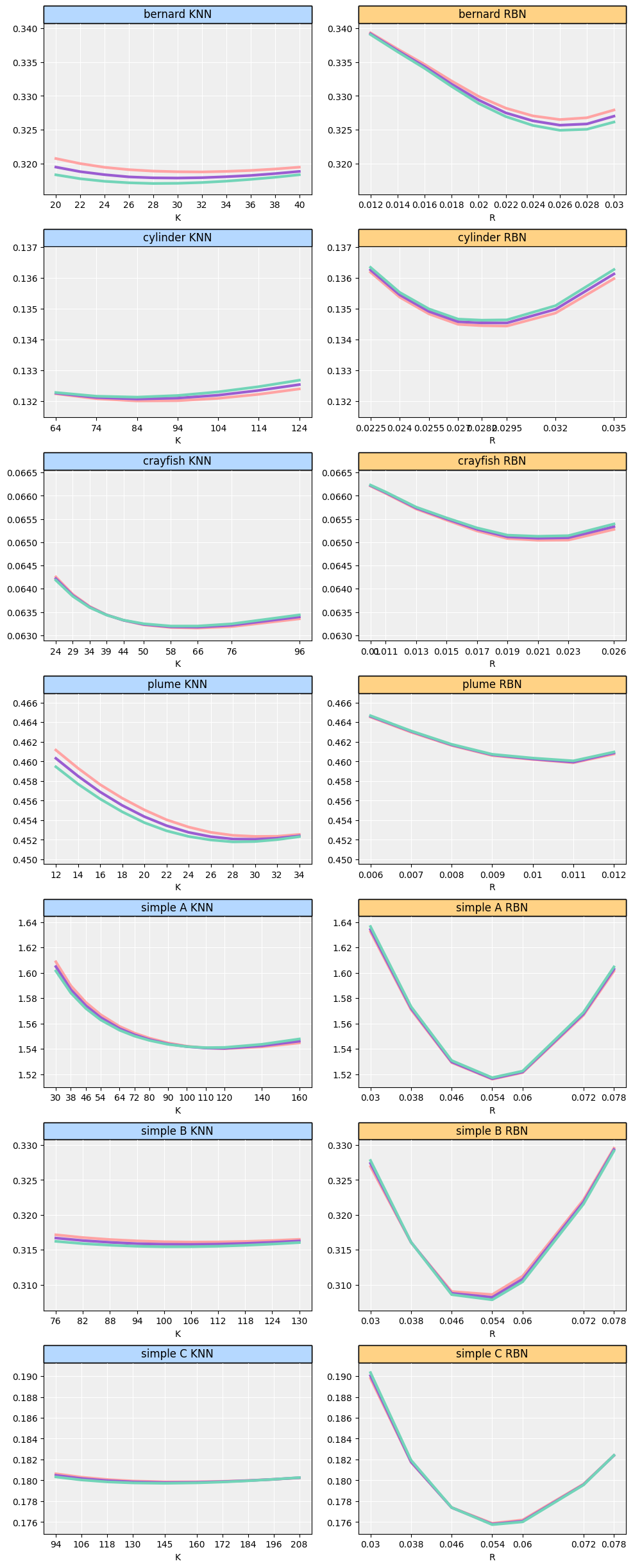}
  \caption{Average error magnitude of the reconstructed vector fields for the seven streamline data sets generated using uniform seeding with KNN (left) and RBN (right), respectively. The three curves shown in each plot correspond to three different distance metrics (i.e., shortest, average, and longest). From these plots, we see that the longest distance metric leads to the smallest errors in more cases than the other two distance metrics, even though their difference is small. Also, KNN outperforms RBN for the real-world simulated flows, while RBN beats KNN for the simple flows. 
  }
  \label{fig:globalObservations}
\end{figure}

\begin{figure}[!t]
\centering
  \includegraphics[width=0.9\columnwidth]{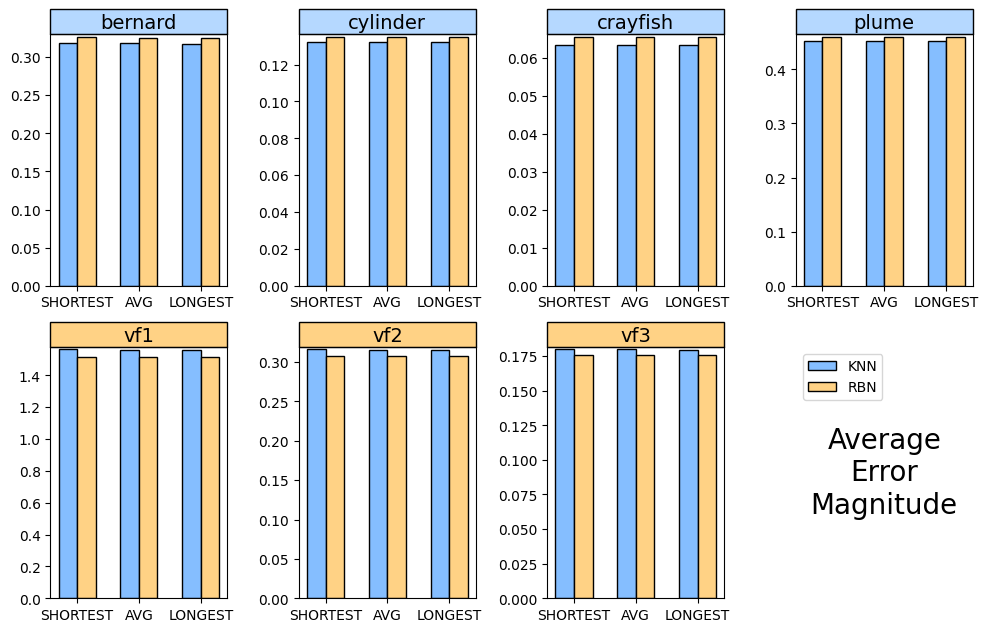}
  \caption{Average error magnitude bar plots of the best reconstruction results (i.e., with the smallest errors) for all seven data sets. Each plot presents three sets, corresponding to the results with the three distance metrics, respectively. Within each set, two bars are shown: one for KNN and one for RBN.}
  \label{fig:errorBarPlots}
\end{figure}


\begin{table}
\centering
\small 
\setlength{\tabcolsep}{1.0pt}
\begin{tabular}{|c|c|c|c|c|c|c|c|c|c|c|c|c|c|}
\hline
\multicolumn{2}{|c|}{bernard} & \multicolumn{2}{c|}{crayfish} & \multicolumn{2}{c|}{cylinder} & \multicolumn{2}{c|}{plume} & \multicolumn{2}{c|}{A} & \multicolumn{2}{c|}{B} & \multicolumn{2}{c|}{C} \\
\hline
 KNN & RBN & KNN & RBN & KNN & RBN & KNN & RBN & KNN & RBN & KNN & RBN & KNN & RBN \\
\hline
\cellcolor{bluecell}L & \cellcolor{bluecell}L & \cellcolor{bluecell}L & \cellcolor{yellowcell}S & \cellcolor{bluecell}L & \cellcolor{yellowcell}S & \cellcolor{bluecell}L & \cellcolor{bluecell}L & \cellcolor{bluecell}L & \cellcolor{yellowcell}S & \cellcolor{bluecell}L & \cellcolor{bluecell}L & \cellcolor{bluecell}L & \cellcolor{bluecell}L \\
\hline
\end{tabular}
\caption{Best-performing distance metric for each data set and method. Here, "L" represents the longest distance metric, while "S" indicates the shortest distance metric.}

\label{tab:distTable}
\end{table}

\begin{figure}[!t]
\centering
  \includegraphics[width=0.9\columnwidth]{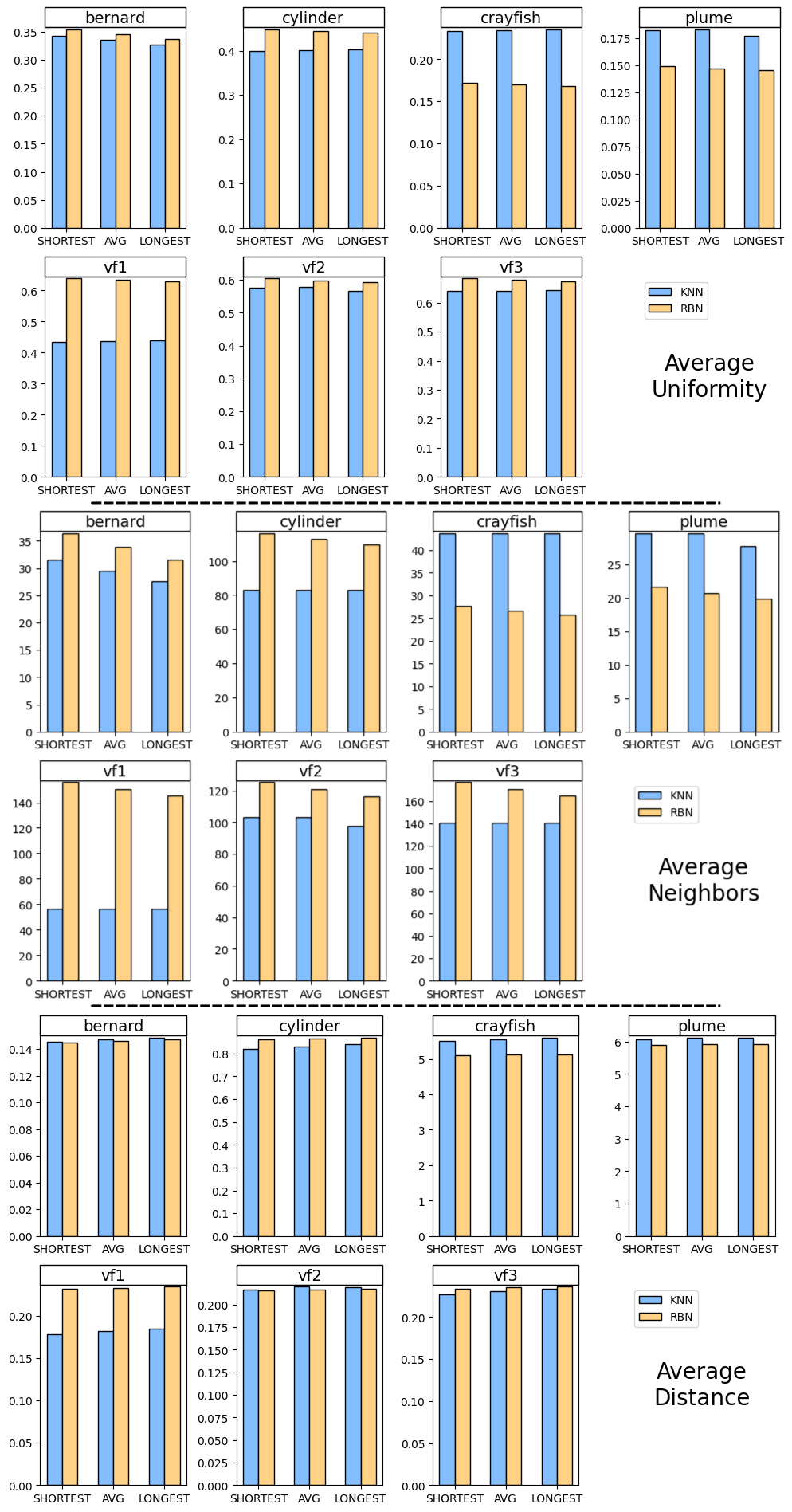}
  \caption{
  Three distinct sets of bar plots, each set corresponding to one of the metrics: average uniformity, average neighbors, and average distances. These plots display the mean values derived from the best reconstruction results—specifically, those with minimal errors—across all seven data sets. For each of the three metrics, the plot is divided into three subsets based on the distance metrics used. Within these subsets, there are two bars juxtaposed: one representing the results for KNN and the other for RBN. This format allows for a straightforward comparison between KNN and RBN across different distance metrics, providing clear insights into their respective performances.}
  \label{fig:miscBarPlots}
\end{figure}

\subsection{Overview of the Performance of Reconstruction Task}
\label{sec:overview_finding}

\autoref{fig:globalObservations} shows the reconstruction error across various \( K \)s and \( R \)s for each data set. All streamline data sets used here were generated using the uniform placement strategy described in Section \ref{sec:setup}. The left column shows the results with KNN while the right column for RBN. For each plot, the three curves correspond to three different distance metrics. In addition to Figure \ref{fig:globalObservations}, we also compare the best KNN result (among all different distance metrics and \( K \)s) with the best RBN result  (among all different distance metrics and \( R \)s)  for each data set in \autoref{fig:errorBarPlots}. From these plots, we see different performances between the KNN and RBN, especially when comparing across different types of flows.  

\begin{itemize}[noitemsep,nolistsep]
    \item \textbf{Complex Flows:} For the four complex flow data sets—Bernard, Cylinder, Crayfish, and Plume—KNN  outperforms RBN, as shown in both \autoref{fig:globalObservations} and \autoref{fig:errorBarPlots}.
    
    \item \textbf{Simple Synthetic Flows:} For the simpler synthetic flows—Simple A, Simple B, and Simple C—RBN shows superior performance compared to KNN.
    
    \item \textbf{Impact of Distance Metrics:} 
    The choice of distance metric did not drastically sway the reconstruction errors. Nonetheless, of the 14 best reconstruction results, spanning 7 data sets with both KNN and RBN methods, $d_{max}$ (longest distance) metric emerges superior \textbf{11 times}, outperforming $d_{min}$ (shortest distance) metric which only managed to do so 3 times (Figure \ref{tab:distTable}). This can be attributed to the fact that employing $d_{max}$ in neighbor search typically yields a segment whose points are closer to the query point (see \autoref{fig:pointlinecase} for an illustration).
    For RBN, $d_{max}$ also leads to reduced numbers of neighbors, which not only helps reduce the computation cost but also the overfitting issue (i.e., including many segments that only have a small portion falling inside the search range). 
    We also wish to point out that computing $d_{max}$ is as fast as computing $d_{min}$.
\end{itemize}



The above overarching findings can be observed across all the evaluated streamline data sets, encompassing the uniformly-seeded,  feature-aware, and randomly seeded streamlines. For a comprehensive exploration of these results, readers are directed to the provided supplemental material. In the following, we mainly focus on the results using streamlines with uniform seeding.
To gain insight into the different performances of KNN and RBN across our data sets, we conducted a deeper examination. 


\subsection{A Holistic View}
\label{sec:simpleflows}

\paragraph{Simple Flows}
For the three simple flows, KNN consistently lagged behind RBN. 
We find that this may be due to the sparse distribution of the streamline segments near the domain boundaries and corners. Since KNN aims to identify K nearest neighbors from the query point, for a large K value, KNN could search too far away from the query point in a sparse region. This leads to segments that cannot accurately represent the flow behavior at the query point.
\autoref{fig:volrendC} (left) highlights large KNN errors near the corners compared to RBN. 
When excluding regions with a sparse streamline placement, we see the average error decrease (Figure \ref{fig:volrendC} (right)). However, even excluding these regions, RBN still slightly outperforms KNN. This suggests that the different neighbor segments returned by KNN are not ideal when compared to those returned by RBN in other regions. This requires a more in-depth inspection, which we will detail in \autoref{sec:good_vs_bad}.


When examining the average properties of the neighbors returned by KNN and RBN for simple flows (Figure \ref{fig:miscBarPlots}), a couple of insights emerge. Firstly, the average neighborhood distance within KNN neighborhoods usually exceeds that within RBN neighborhoods. In contrast, the average uniformity values for KNN neighborhoods are generally lower than those for RBN neighborhoods. This could be because KNN has trouble locating neighbors sufficiently close to the query point near the corners and boundaries of those flows as already illustrated in Figure \ref{fig:volrendC} in the Appendix. Meanwhile, since RBN neighborhoods have smaller average neighbor distances and larger uniformity than KNN neighborhoods, we expect RBN to result in smaller reconstruction errors.



\paragraph{Complex Flows}
For the four complex flows, KNN outperforms RBN. 
Most of these flows contain vortices with different scales where streamlines exhibit strong rotational behavior, leading to a denser distribution. 
Given a grid point in one of these vortex regions, KNN will likely identify the K nearest neighbors that are close to the grid point. In contrast, RBN will identify all neighbors within the search sphere with a fixed radius. Due to the quick change of the flow direction in these regions, the segments near the boundary of the search sphere may have very different orientations from those closer to the query grid point. Including these 'farther' segments for the reconstruction of the vector at the grid point will likely increase the error. This behavior of RBN shares some similarities to the overfitting issue. 
Nonetheless, KNN still suffers from a similar issue as in the simple flows in the regions with a sparse placement of streamlines. 


In \autoref{fig:miscBarPlots}, an intriguing pattern emerges for the Plume, Crayfish, and to a lesser extent, the Bernard data sets. \emph{Despite KNN registering a higher global average distance than RBN, it consistently outperforms RBN in terms of lower average error magnitudes}. This counter-intuitive observation can be attributed to the presence of sparse RBN regions, as discussed in \autoref{sec:KNNorRBN}. These specific data sets possess an abundance of grid points where RBN returns only a single neighbor. While such a configuration guarantees a lower average distance (since there's only one neighbor to account for), it often translates to a higher reconstruction error. This is because a single neighbor, while close to the query point, might not sufficiently represent the local behavior of the vector field, making it less robust than a KNN configuration that incorporates multiple neighbors. Therefore, data sets characterized by numerous such regions tend to exhibit deceivingly lower global average distances for RBN. However, the higher average errors in these areas clearly highlight the limitations of depending exclusively on RBN in such situations.

However, this does not mean KNN is always better than RBN. The fact that KNN achieves optimal results with fewer average neighbors might also indicate that at higher K values, reconstruction quality decreases for various reasons. This phenomenon is particularly evident in simple flows where RBN outperforms KNN. Furthermore, the diminished average spatial uniformity in KNN may result in an inadequate representation of the local flow behavior within the neighborhood.

From the above inspection of the overall (average) characteristics of all the neighborhoods, we see that \textbf{the characteristics of the KNN and RBN neighborhoods do not always match the common intuition}. To better understand this discrepancy, we now subdivide the grid points of each flow into subgroups and look at the characteristics of their respective neighborhoods.



\subsection{Delineating ``Good'' and ``Bad'' Neighborhoods}
\label{sec:good_vs_bad}





We categorically divide grid points based on their respective reconstruction errors. For each data set, we obtain the two most exemplary reconstruction results: one is obtained using KNN and the other with RBN. From each of these outcomes, we curate two distinct groups. The ``Good'' neighborhood group comprises neighborhoods with reconstruction errors in the first 10th percentile (i.e., the first 10\% \footnote{We selected this percentile to better reveal the difference between the two groups. The usual 25\% percentile also reveals similar behavior.} of neighborhoods with the smallest errors). In contrast, the ``Bad'' group encompasses neighborhoods with errors at and above the 90th quartile (i.e., the last 10\% of neighborhoods with the largest errors). This stratification ensures an equivalent count of neighborhoods across both categories for a statistically fair comparison.

\begin{figure}[!t]
\centering
  \includegraphics[width=0.95\columnwidth]{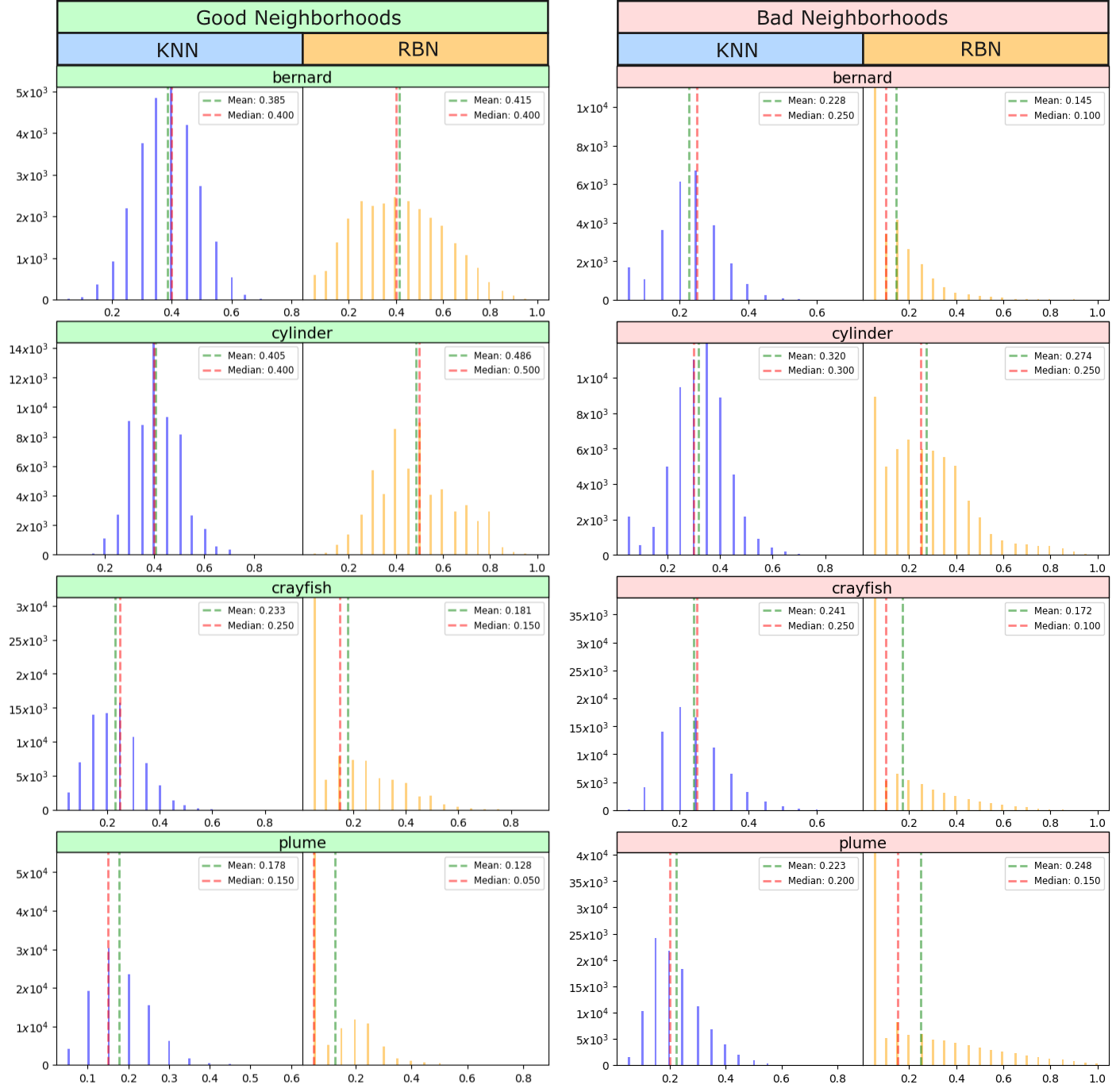}
  \caption{Distributions of uniformity for the ''Good'' (left) and ``Bad'' (right) neighborhoods in the complex flows.}
  \label{fig:complexHistUni}
\end{figure}

\begin{figure}[!t]
\centering
  \includegraphics[width=0.95\columnwidth]{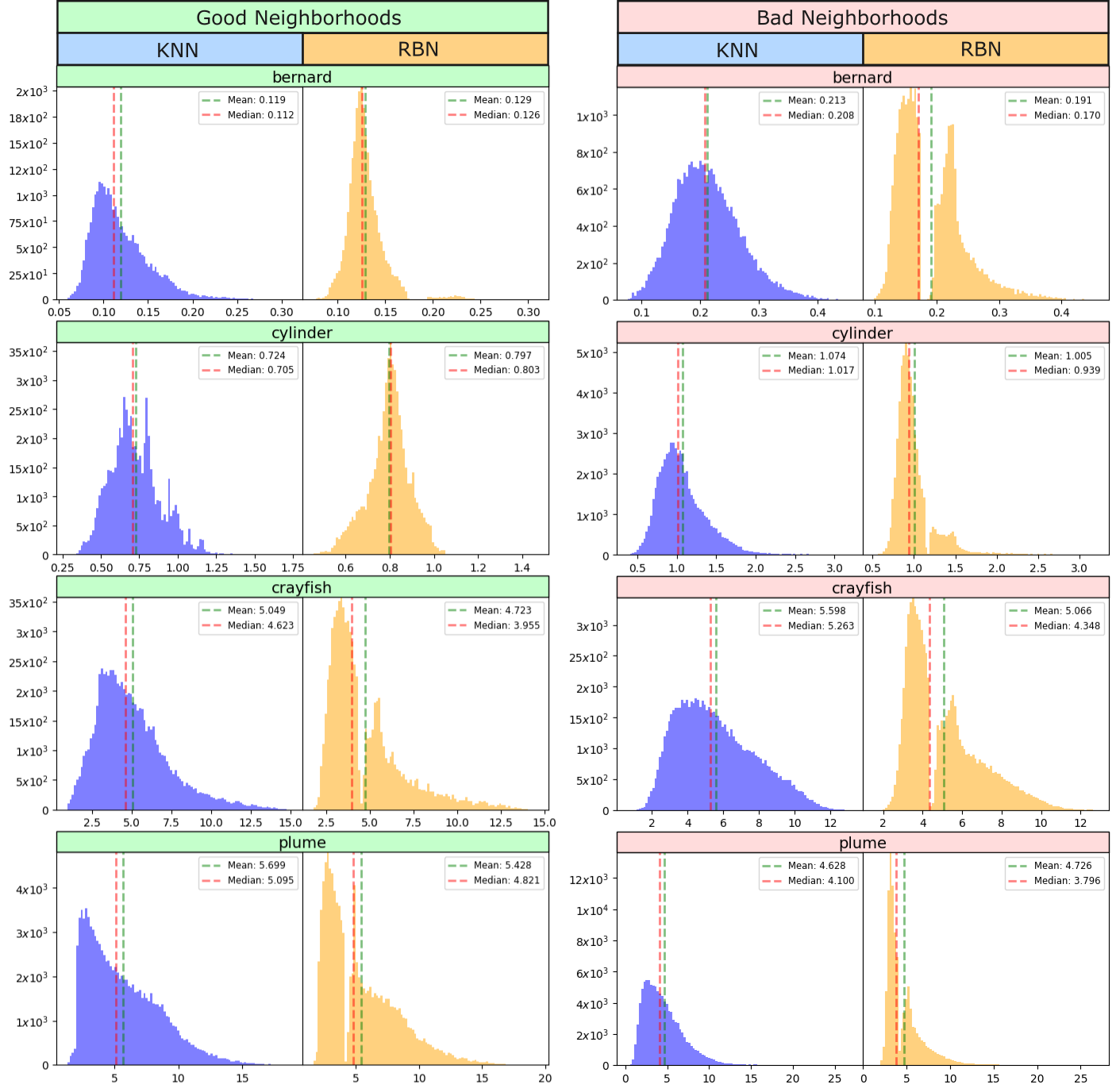}
  \caption{Distributions of average distances for the ``Good'' (left) and ``Bad'' (right) complex flow neighborhoods.}
  \label{fig:complexHistDist}
\end{figure}

Next, we perform a comparative analysis of these ``Good'' and ``Bad'' groups for both KNN and RBN across all seven data sets. This involves charting histograms of the average attributes of the selected neighborhoods, including average neighbor distance (to the query point) and uniformity. The intent is to discern any consistent traits or patterns intrinsic to each group.  \autoref{fig:histSimpleUni},\autoref{fig:histSimpleDist}, \autoref{fig:complexHistUni}, and \autoref{fig:complexHistDist} show the histograms of these attributes for the simple and complex flows, respectively. 

In examining the inherent characteristics of ``Good'' and ``Bad'' neighborhoods, we discern several distinguishable patterns. For example, ``Bad'' neighborhoods typically have greater average distances and smaller uniformity.
In contrast, the ``Good'' neighborhoods exhibit smaller average distances and larger uniformity.
However, two complex flow data sets exhibit a different behavior from the others.
Also, the value ranges of those attributes for the ``Good'' and ``Bad'' neighborhoods have significant overlap. 
This makes a clear cut of the value ranges of the involved attributes (e.g., $(0, 0.3)$ for the average neighbor distance) that may correspond to ``Good'' or ``Bad'' neighborhoods difficult. In the following, we attempt to provide some explanation of the different performances between KNN and RBN in the ``Good'' and ``Bad'' neighborhoods of the simple and complex flows, respectively.

\subsubsection{Simple Flow Comparison}
\begin{figure}[!thb]
\centering
  \includegraphics[width=0.95\columnwidth]{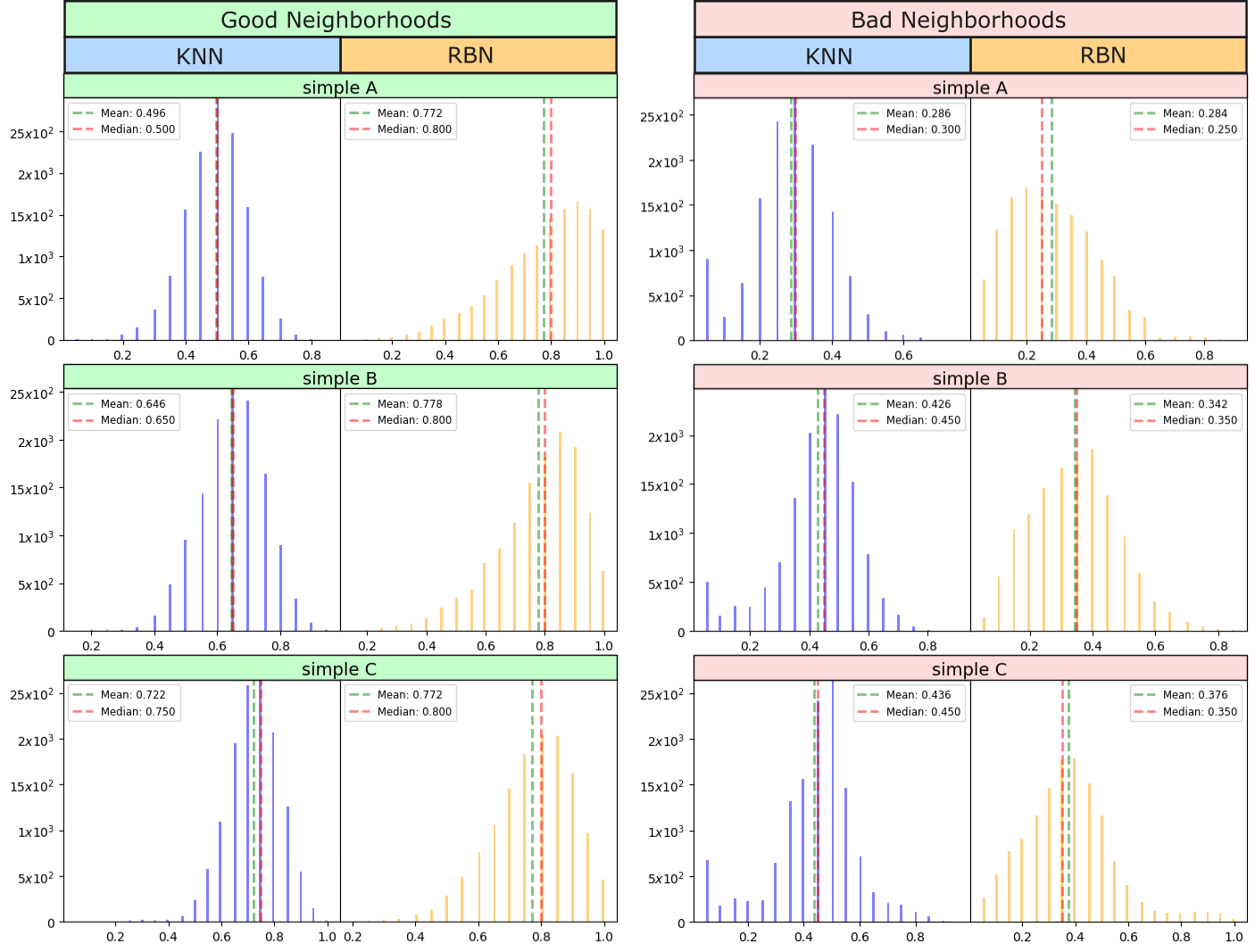}
  \caption{Uniformity distributions of the ``Good'' and ``Bad'' neighborhoods in the simple flow data sets. ``Good'' neighborhoods have much higher uniformity values than those of ``Bad'' neighborhoods.
  }
  \label{fig:histSimpleUni}
\end{figure}

\begin{figure}[!thb]
\centering
  \includegraphics[width=0.95\columnwidth]{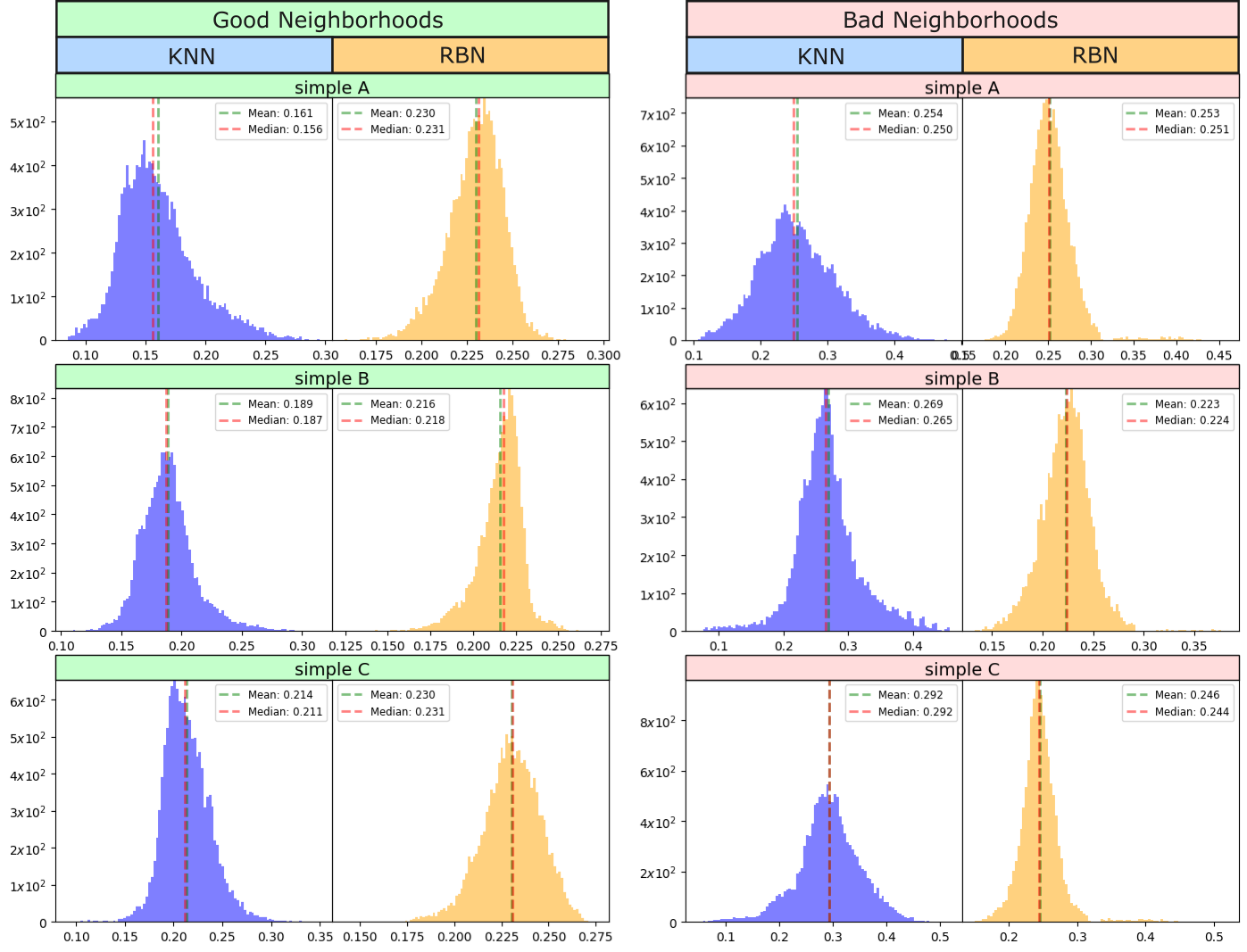}
  \caption{Distributions of average neighbor distances for the ``Good'' and ``Bad'' simple flow neighborhoods. ``Good'' neighborhoods have smaller average distances than those of ``Bad'' neighborhoods.
  }
  \label{fig:histSimpleDist}
\end{figure}

\autoref{fig:histSimpleUni} and \autoref{fig:histSimpleDist} show what we expect for the ``Good'' and ``Bad'' neighborhoods. That is, ``Good'' neighborhoods (left column) tend to have smaller average neighbor distances (\autoref{fig:histSimpleDist}) and larger uniformity values (\autoref{fig:histSimpleUni}) than those of ``Bad'' neighborhoods (right column). This is true for both KNN and RBN results. In the meantime, while KNN usually leads to smaller average distances than RBN for the ``Good'' neighborhoods, they tend to have larger average distances than RBN for the ``Bad'' neighborhoods. This again could be because KNN is struggling to identify close-enough neighbors in the sparse regions. On the other hand, the uniformity values of RBN results in the ``Good'' neighborhoods are larger (better) than those of KNN, while they are worse in the ``Bad'' neighborhoods. Since the average distances do not vary much between the ``Good'' and ``Bad'' neighborhoods for the RBN results, the large differences in the uniformity values between the ``Good'' and ``Bad'' neighborhoods likely impact the reconstruction errors with the RBN. In contrast, the reconstruction errors seem to depend on both the average distance and uniformity of the neighbors returned by KNN in these simple flows.




\label{complexcompare}

\subsubsection{Complex Flows Comparison}
\label{sec:complexflows}
Unlike the simple flows, the ``Good'' and ``Bad'' neighborhoods exhibit inconsistent behaviors. For the uniformity values (\autoref{fig:complexHistUni}), the neighborhoods of Bernard and Cylinder data sets have the expected behavior (i.e., ``Good'' neighborhoods have larger uniformity than the ``Bad'' neighborhoods). However, an opposite behavior is observed for the Crayfish and Plume data sets (i.e.,  ``Good'' neighborhoods have smaller uniformity than the ``Bad'' neighborhoods). For the average neighbor distances (\autoref{fig:complexHistDist}), the neighborhoods of Bernard,  Cylinder, and Crayfish data sets exhibit the expected behavior (i.e., ``Good'' neighborhoods have smaller average distances than the ``Bad'' neighborhoods), while the neighborhoods of the Plume data set have an opposite behavior (i.e., ``Good'' neighborhoods have larger average distances than the ``Bad'' neighborhoods).


It is worth mentioning that the histograms for RBN's average distance exhibit a bimodal pattern, different from the unimodal distribution exhibited by KNN. 
This bimodal behavior of RBN's average distance distribution is attributed to how our RBN is implemented (\autoref{sec:pbneighborsearcn}). Recall that in our modified RBN approach, if within a given search range determined by R, no segments are identified, we use the closest neighbor whose distance to the query point is larger than R; otherwise,  at least one neighbor with a distance less than R is found. This divides the RBN neighborhoods into two groups (i.e., neighborhoods with average neighbor distance > R and those with average distance < R).

Now, to understand why Crayfish and Plume data sets have the opposite behavior to the others, we look into the configuration of their input streamlines. We found that each of these two data sets has a large portion of the regions with low streamline distribution (i.e., sparse) even with a uniform seeding strategy. Specifically, 56\% of the RBN neighborhoods in Plume contain just one neighbor (see \autoref{fig:pluVolRend} in the Appendix), and it is 36\% in Crayfish. 
This could be caused by the low-velocity magnitude there so the streamline may not go far (i.e., similar to reaching a critical point). Note that approximately 20\% of its grid points in Crayfish exhibit a magnitude of zero.
To see whether this sparsity causes the opposite behavior of the two data sets to the others', we exclude those grid points whose RBN neighborhoods have only 1 neighbor. We then analyze the histograms of uniformity and average distance of the remaining neighborhoods and find:

\begin{enumerate} [noitemsep,nolistsep]
    \item For the remaining neighborhoods of Crayfish, the ``Bad'' neighborhoods exhibit reduced uniformity compared to the ``Good'' neighborhoods.
    \item For the remaining neighborhoods of Plume, the ``Bad'' neighborhoods now have a larger average distance compared to the ``Good'' neighborhoods. However, the ``Bad'' neighborhoods still have slightly larger uniformity values than the ``Good''  neighborhoods.
\end{enumerate}

This additional analysis shows how the sparsity in the streamline distribution may impact the neighbor search. In short, \emph{in the sparse areas of the input streamlines, the common intuition of an ideal neighborhood with neighbors close to and uniformly distributed around the query point does not always apply}. Unfortunately, these sparse regions often accompany low-velocity magnitude that often exists in different flows. To address this, a modified criterion for ``good'' neighborhoods in sparse areas is needed.

\subsubsection{KNN and RBN Similar vs Dissimilar Neighborhoods}
\label{inclusivevsexclusive}
\begin{figure}[!t]
\centering
  \includegraphics[width=0.95\columnwidth]{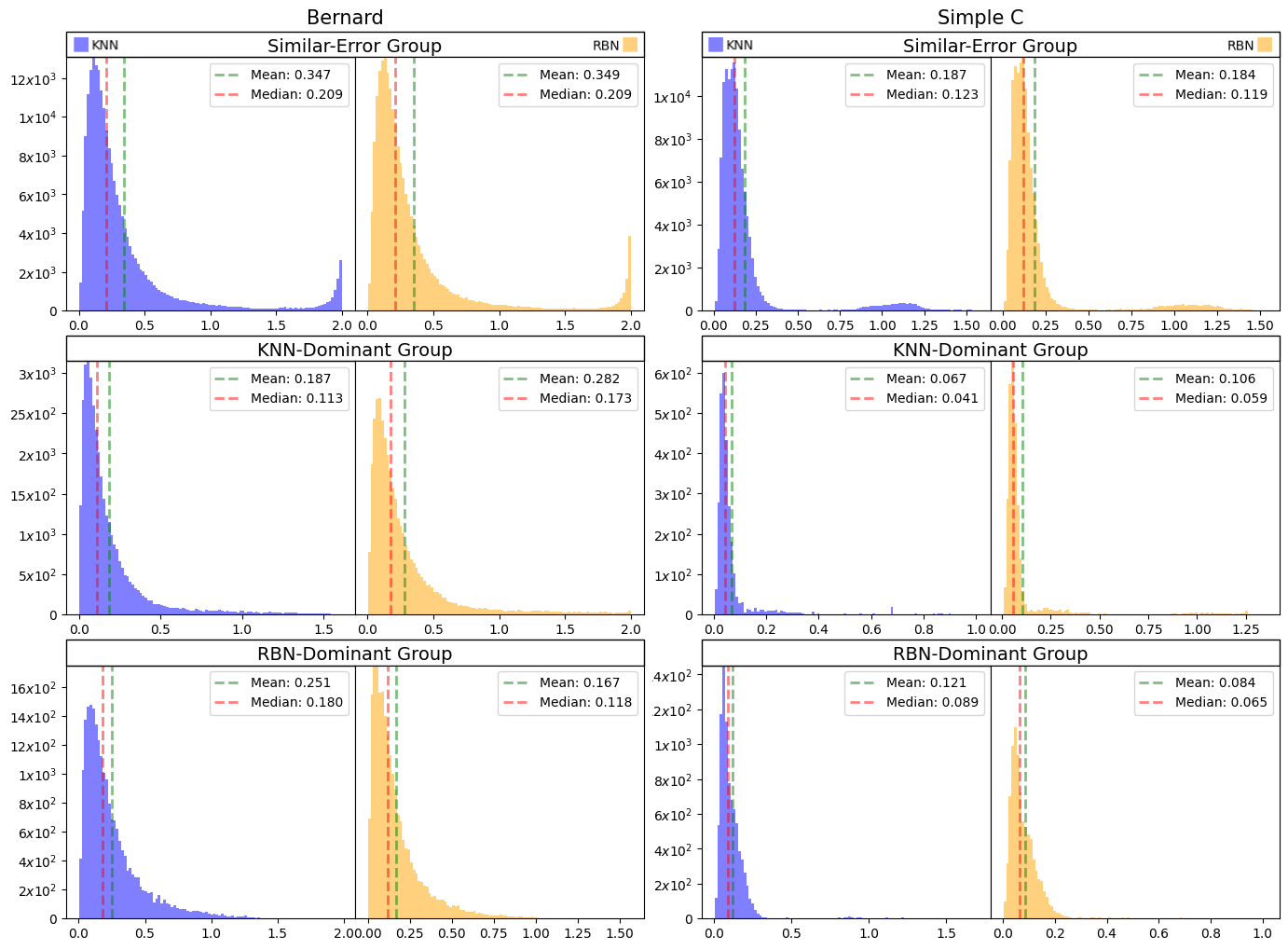}
  \caption{Comparative analysis of gridpoints across the Bernard and Simple B data sets, categorized into three distinct groups: Similar-Error, KNN-Dominant, and RBN-Dominant. The distribution patterns showcase the relative performance of KNN and RBN for each data set, with the Bernard data set displaying a predominant dominance of KNN, while the Simple C data set leans towards RBN's efficacy.}
  \label{fig:InclusiveExclusive}
\end{figure}

We now provide an additional comparison between KNN and RBN. To do so, we divide the grid points into three groups.

\begin{enumerate} [noitemsep,nolistsep]
    \item Similar-Error Group: Gridpoints with the differences of the errors of both KNN and RBN are small.
    \item KNN-Dominant Group: Gridpoints where KNN has much smaller errors than RBN.
    \item RBN-Dominant Group: Gridpoints where RBN has much smaller errors than KNN.
\end{enumerate}

To determine the grouping, we compute the absolute percentage difference at each grid point defined below.
$$
\text{Percentage Difference} = \frac{\left| \text{Error}_{\text{KNN}} - \text{Error}_{\text{RBN}} \right|}{\min(\text{Error}_{\text{KNN}}, \text{Error}_{\text{RBN}})} \times 100
$$

We empirically selected a threshold at the 80th percentile using the above error difference, meaning that 80\% of all gridpoints, when sorted by percentage difference, lie to the left of this threshold. They are considered in the Similar-Error Group.

\autoref{fig:InclusiveExclusive} provides a comparative analysis of these three groups across the Bernard and Simple C data sets. It shows the following.

\begin{itemize} [noitemsep,nolistsep]
    \item For the Bernard data set, KNN consistently achieves a lower average error relative to RBN. Consequently, the KNN-Dominant group more than doubled the RBN-Dominant Group.
    \item Conversely, the Simple B data set, which has a more favorable error rate for RBN, contains a larger number of gridpoints in the RBN-Dominant group compared to the KNN-Dominant group.
    \item On closely inspecting the error differences between KNN and RBN within the same groups, it is evident that their differences are minimal for most cases. However, in the Bernard data set, the discrepancy between KNN and RBN errors for the two dominant groups is more than double, suggesting notable variances in their performances. Simple C doesn't exhibit as pronounced a gap, which could be attributed to the data set's predictable flow patterns that both KNN and RBN manage to capture efficiently.
\end{itemize}


\subsection{KNN or RBN for Vector Field Reconstruction}
\label{sec:KNNorRBN}

\begin{figure}[!t]
\centering
  \includegraphics[width=0.5\columnwidth]{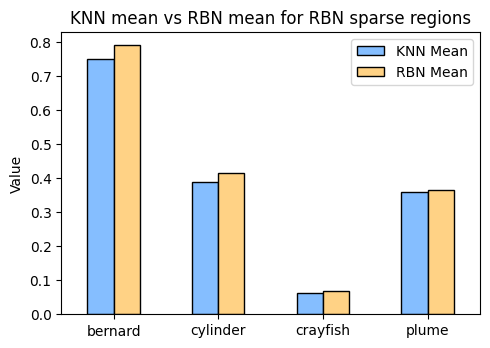}
  \caption{Average Error Comparison between KNN and RBN for sparse RBN regions, where RBN defaults to K=1 due to the absence of neighbors within a set radius, R.}
  \label{fig:RBNSparse}
\end{figure}


The above analysis aims to identify the distinct attributes of ``Good'' and ``Bad'' neighborhoods derived from both the KNN and RBN methods for the vector field reconstruction task. 
Unfortunately, due to the large overlap between the histogram distributions of the average neighbor distance and uniformity measures for the ``Good'' and ``Bad'' neighborhoods, a guideline based on separate and distinct characteristics of these distributions is challenging. 
In fact, we have tried to classify individual neighborhoods into ``Good'' and ``Bad'' groups using their respective attributes, but were unable to achieve this. Nevertheless, we have derived a few insights for the neighbor selection for vector field reconstruction task, which are summarized below.

In general, neighbors that are close to and evenly distributed around the query point are still preferred. In practice, this is challenging to achieve and highly depending on the configurations of the input streamlines. 
Between KNN and RBN, KNN should be considered first to handle any input streamline sets. This is particularly important for sparse areas. \autoref{fig:RBNSparse} shows the average error comparison between KNN and RBN in sparse areas where RBN can only find one neighbor. In those areas, RBN tends to have a larger error than KNN that finds more neighbors. This strongly suggests that in sparse areas when the closest neighbors to a query point are further away, more than one neighbor usually performs better. This further suggests the use of KNN there. However, we should be aware that KNN with a larger K may find neighbors that are too far away from the query to be useful. Therefore, an adaptive strategy needs to be developed to handle the sparse regions. 
Furthermore, the longest distance metric should be considered to work with a neighbor search algorithm to select closer neighbors to the query point. 

Our findings can potentially be used to improve the current practice of neighbor-based vector field reconstruction approach. For example, one can identify four closest neighbors to the query that form a tetrahedron enclosing the query point. Then, a barycentric interpolation can be performed to approximate the vector value at the query point from the four closest neighbors. 
In the meantime, our findings show that more neighbors need not lead to more accurate reconstructed vectors. This strongly suggests that a subset of the identified neighbors (whether there are four or not) can be applied to reconstruct the vector at a query point with a lower error. To improve the current practice, one can use the proposed measures to select the four neighbors that are more evenly distributed around the query point while maintaining a small average distance to the query point.

Finally, a new neighbor search method may be needed for the vector field reconstruction task that can adapt to the varying configurations of the input streamlines.

\section{Point-Saliency Prediction Results}
\label{subsec:saliencyresults}

We evaluated the performance of KNN and RBN in predicting point-saliency across our seven datasets using both the Concordance Correlation Coefficient (CCC) and Pearson Correlation Coefficient (PCC). The results reveal several interesting patterns and differences between the two neighbor search strategies.

\subsection{CCC and PCC Analysis}
\label{subsubsec:cccpccanalysis}

\begin{figure}[!t]
\centering
\includegraphics[width=\linewidth]{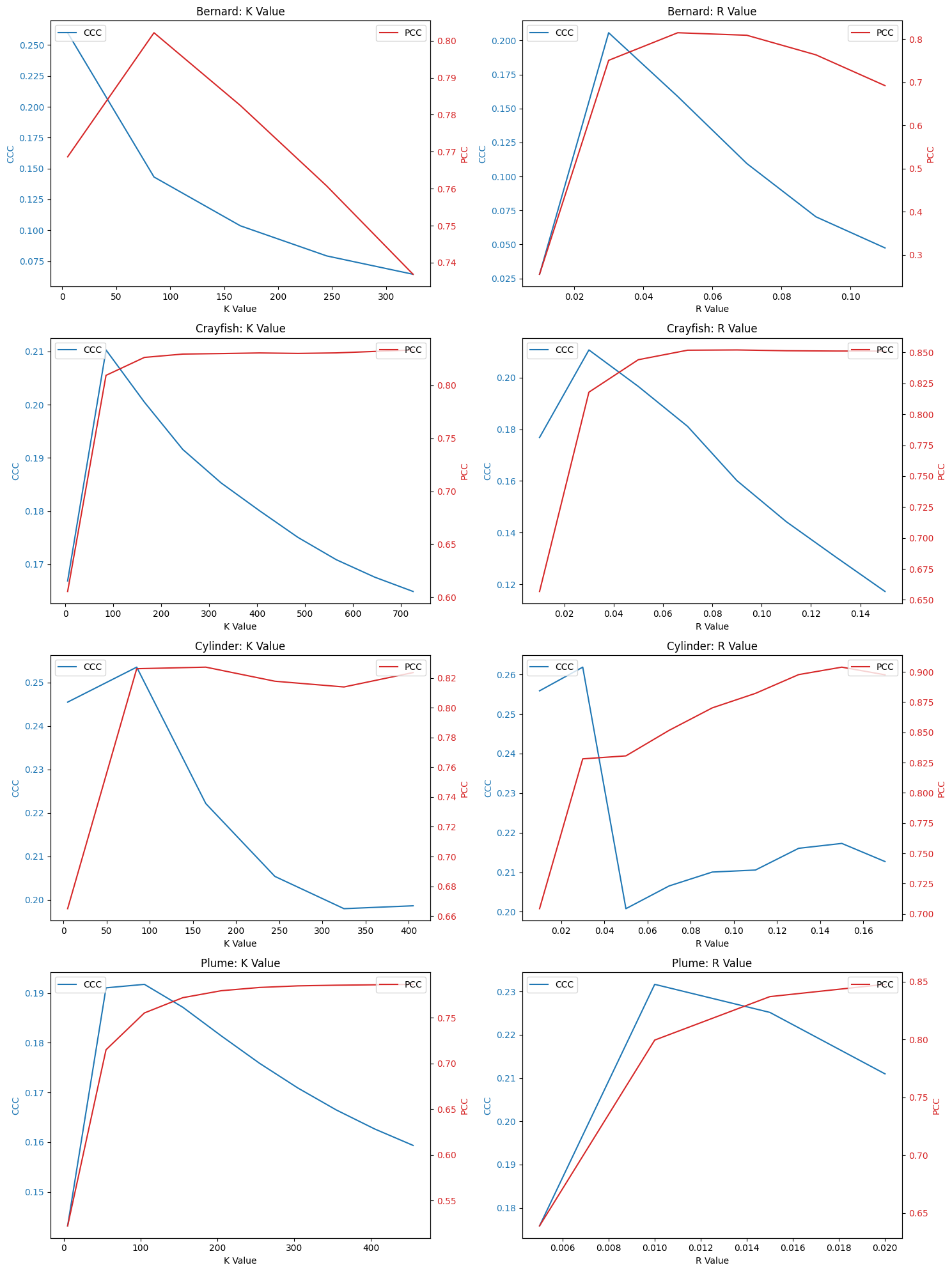}
\caption{CCC and PCC vs. K and R Values for five complex Flow Datasets}
\label{fig:ccc_pcc_plots}
\end{figure}

Figure \ref{fig:ccc_pcc_plots} shows the CCC and PCC values for different K (KNN) and R (RBN) values across five complex flow datasets. The CCC metric (blue line) generally shows a clear convergence to optimal K and R values for both KNN and RBN across most datasets. This convergence indicates the existence of an optimal neighborhood size for saliency prediction, beyond which the prediction quality deteriorates.

Interestingly, the Bernard dataset exhibits an anomalous behavior for KNN, with CCC values decreasing monotonically from K=2. This unique pattern suggests that for this particular flow, even small increases in the number of neighbors lead to less accurate saliency predictions, possibly due to the complex local flow structures in the Bernard convection.

In contrast to CCC, the PCC metric (red line) shows a different behavior. For most datasets, PCC increases monotonically with K and R, failing to indicate an optimal neighborhood size. The Bernard dataset is an exception, where PCC does converge to an optimal value. This difference between CCC and PCC underscores the limitations of using correlation alone as an evaluation metric. While PCC captures the linear relationship between predicted and reference saliency values, it doesn't account for systematic biases or scale differences, which CCC does consider.

It's worth noting that for all datasets, the p-values for PCC were consistently 0, indicating statistically significant correlations. However, as our results show, statistical significance doesn't necessarily imply practical significance or optimal performance.

\subsection{Visual Comparison of Saliency Predictions}
\label{subsubsec:visualcomparison}

\begin{figure}[!t]
\centering
\includegraphics[width=\linewidth]{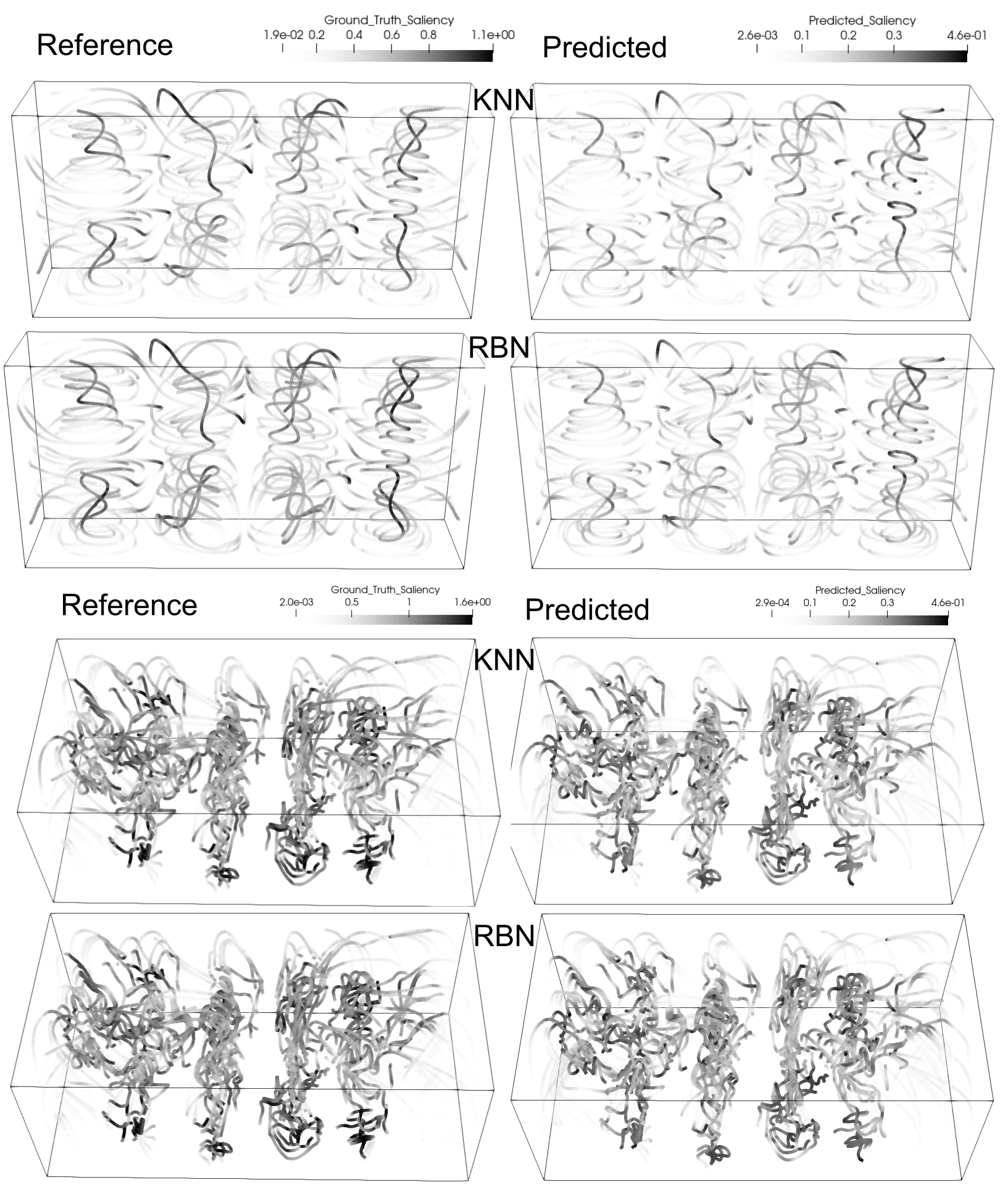}
\caption{Comparison of reference and predicted point saliency for Bernard (top) and Crayfish (bottom) flow datasets using KNN and RBN neighbor search methods. For each dataset, the left column displays the reference saliency, and the right column shows the predicted saliency using KNN (top row) and RBN (bottom row) methods. Saliency values are represented by grayscale intensity, with darker regions indicating higher saliency.}
\label{fig:saliency_comparison}
\end{figure}

Figure \ref{fig:saliency_comparison} presents a side-by-side visual comparison of the best KNN and RBN saliency predictions alongside the reference saliency for the Bernard and Crayfish datasets. These visualizations reveal notable differences between the two neighbor search strategies.

Across both datasets, the KNN predicted saliency maps appear more "faded" compared to their RBN counterparts, particularly in regions with sparse streamline coverage. This visual difference suggests that KNN may struggle to produce sufficiently high saliency values in areas where neighboring points are likely to come from the same or similar streamlines. In contrast, RBN seems less affected by this issue, potentially due to its ability to capture a more diverse set of neighbors in sparse regions.

The reference saliency maps, as expected, show the most detailed and pronounced patterns of flow complexity. Both KNN and RBN predictions capture the overall structure of these patterns, but with varying degrees of intensity and detail. RBN results show higher contrast and more defined salient features, suggesting that RBN may be more effective at identifying areas of high saliency in these flow fields.

These visual comparisons align with our quantitative findings, reinforcing the idea that the choice of neighbor search strategy can significantly impact the quality of saliency predictions, especially in regions with varying streamline density.

\subsection{Error Analysis}
\label{subsubsec:erroranalysis}

While not visualized here due to space constraints, our analysis of raw error values (available in the supplemental material) reveals a consistent increase in error as K and R increase. However, this trend is primarily due to the changing reference sample radius rather than a degradation in prediction quality. As K or R increases, the average distance to neighbors grows, necessitating a larger radius for sampling the reference saliency. This expanding radius introduces additional flow complexity that the streamline-based predictions struggle to capture fully, leading to increased errors.

This observation highlights the challenge of directly comparing error values across different K and R settings and further justifies our use of CCC as the primary evaluation metric.

In conclusion, our saliency prediction results provide valuable insights into the strengths and limitations of KNN and RBN for capturing local flow characteristics. These findings complement our vector field reconstruction results, offering a more comprehensive understanding of how different neighbor search strategies perform in various aspects of curve-based vector field processing.
\section{Summary and Future Work}
\label{sec:conclusion}

In this work, we present a comprehensive study on how different neighbor search strategies, i.e., KNN and RBN, combined with various distance metrics, impact the neighborhood search for the processing of curve-based vector field data. 
We conduct a large number of tests with different combinations of search strategies and distance metrics, as well as varying parameter setups, using different streamline data sets derived from seven flow data sets. 
These tests yield important observations (Section \ref{sec:results}).
To understand these observations, we propose a few measures to characterize the configurations of neighboring segments, including a novel uniformity measure. 
These measures allow us to associate different configurations with the accuracy and performance of vector field reconstruction results for different search strategy combinations. 
With these measures and the obtained observations, we attempt to introduce a simple guideline for selecting proper neighbors for the vector field reconstruction tasks. Our assessment shows the effectiveness of this guideline. Future endeavors will delve deeper, refining these guidelines and potentially revealing more nuanced insights. For now, these preliminary findings pave the way for researchers navigating the labyrinth of available configurations, aiming to optimize vector field evaluations.
\paragraph{Limitations and future work.} While our study provides the first comprehensive set of information on how existing neighbor search strategies generate different neighbor configurations and how they may impact subsequent processing and analysis tasks, our study still has a few limitations. First, we have made a few simplified assumptions about the input streamlines to facilitate our study, including the use of the numerical integrator with fixed step sizes to compute streamlines and the simple segment decomposition strategy. Second, the measure of the quality of the reconstruction vector field is not comprehensive. In particular, it only considers point-wise errors and does not consider the error between the streamlines extracted from the reconstructed field and the input streamlines. The latter error could provide additional insight. Third, the proposed measures for characterizing the configurations of different neighboring segment configurations are still simple. For example, the proposed uniformity measure is overly simplified and does not accurately depict the evenness of the spatial distribution of the neighbors surrounding a query point. 
Finally, our study only focuses on streamline data sets, while more answers may be sought for the pathline data sets and their processing. In addition, the more important and challenging curve-centered neighbor search problem has not been investigated. 
All these limitations should be properly addressed in future works.

\bibliographystyle{abbrv-doi-hyperref}

\bibliography{Content/references}

\clearpage
\appendix
\section*{Appendix}

\begin{figure}[!h]
\centering
  \includegraphics[width=0.92\columnwidth]{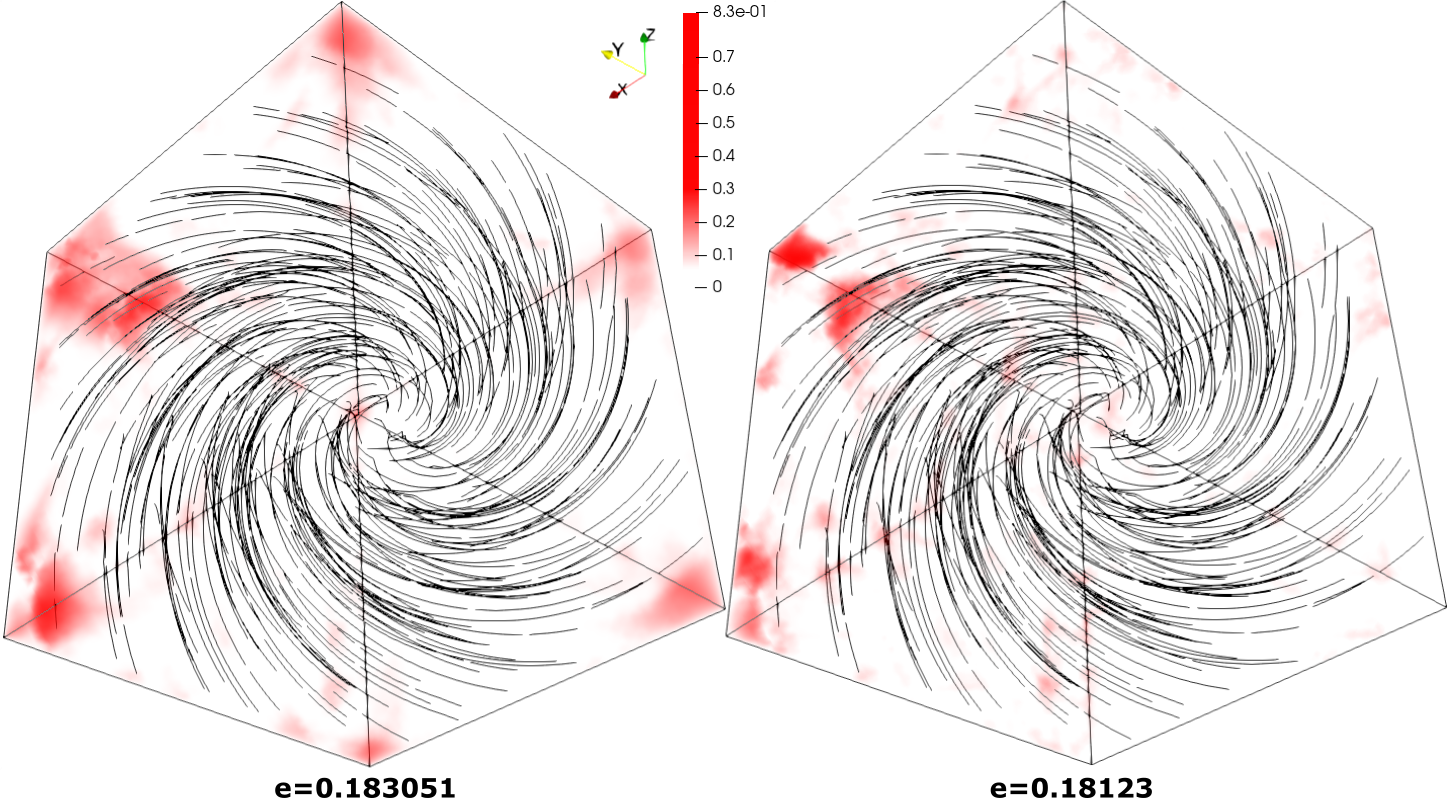}
  \caption{Volume renderings contrasting reconstruction outcomes for the Simple C flow. The best KNN and RBN results are juxtaposed. Regions tinted red signify areas where RBN excelled over KNN. The left depicts the most optimal KNN result, whereas the right displays the same KNN result, albeit with the exclusion of neighbors exceeding a maximum radius \( R=0.4 \). The average error correspondingly diminished from \( 0.183 \) (left) to \( 0.181 \) (right). }
  \label{fig:volrendC}
\end{figure}

\begin{figure}[!h]
\centering
  \includegraphics[width=0.65\columnwidth]{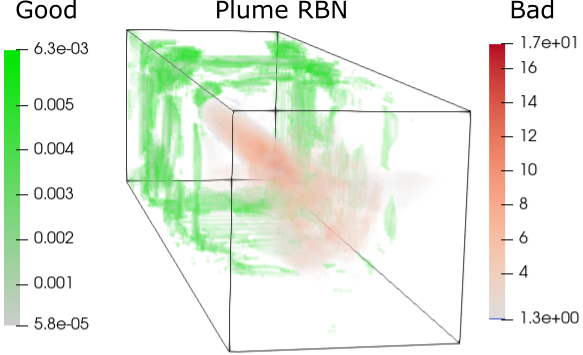}
  \caption{Volume render comparison of the Plume data set. The green regions represent the "Good" neighborhoods and illustrate areas near the outer boundary of the Plume data set. In contrast, the red regions represent the "Bad" neighborhoods and are located much closer to the main body of the Plume.}
  \label{fig:pluVolRend}
\end{figure}

Figure \ref{fig:pluVolRend} offers further insight into the sparsity issue described in Section \ref{sec:complexflows}. It shows that the ``Good'' neighborhoods, which are marked by smaller errors, are primarily located in the sparse areas near the periphery of the Plume data set. In contrast, the ``Bad'' neighborhoods concentrate around the dense core of Plume's primary structure. As a result, despite being located in more peripheral regions, the ``Good'' neighborhoods have greater average distances compared to the ``Bad'' ones.

\begin{figure}[!h]
\centering
  \includegraphics[width=0.55\columnwidth]{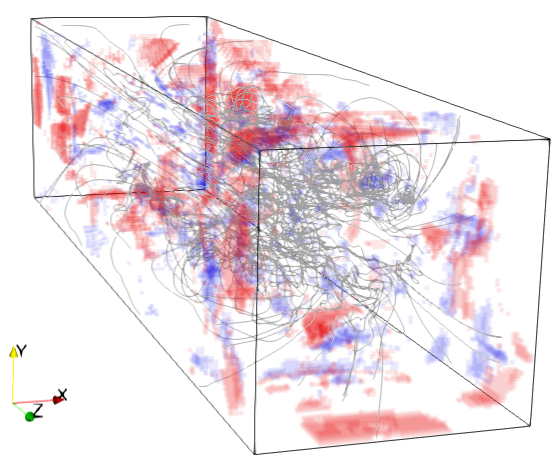}
  \caption{Volume render comparison of the Plume data set. The Blue regions represent the KNN-Dominant neighborhoods and the red regions represent the RBN-Dominant neighborhoods.}
  \label{fig:KRDominant}
\end{figure}


\end{document}